\documentclass[a4paper, accepted=2023-10-23]{quantumarticle}
\pdfoutput=1
\usepackage[utf8]{inputenc}
\usepackage[english]{babel}
\usepackage[T1]{fontenc}
\usepackage{amsmath}
\usepackage[sort&compress,numbers]{natbib}
\usepackage{hyperref}

\usepackage{amsthm}
\usepackage{mathtools}
\usepackage{amsfonts}
\usepackage{amssymb}
\usepackage{bbm}
\usepackage{listings}
\usepackage[usenames,dvipsnames]{xcolor}
\usepackage{algorithmicx}
\usepackage{algorithm}
\usepackage{algpseudocode}
\usepackage{setspace}
\usepackage{braket}
\usepackage{graphicx}

\usepackage[caption=false]{subfig}
\renewcommand{\vec}[1]{\boldsymbol{\mathbf{#1}}}
\DeclareMathOperator*{\argmin}{arg\,min}

\newcommand{\appropto}{\mathrel{\vcenter{
			\offinterlineskip\halign{\hfil$##$\cr
				\propto\cr\noalign{\kern2pt}\sim\cr\noalign{\kern-2pt}}}}}

\begin{document}
	
      \title{Robust Extraction of Thermal Observables from State Sampling and Real-Time Dynamics  on Quantum Computers}

	\author{Khaldoon Ghanem}
	\author{Alexander Schuckert}
	\author{Henrik Dreyer}
	\affiliation{Quantinuum, Leopoldstrasse 180, 80804 Munich, Germany}

%	\date{\today}
	
	\begin{abstract}
	Simulating properties of quantum materials is one of the most promising applications of quantum computation, both near- and long-term. 
	While real-time dynamics can be straightforwardly implemented, the finite temperature ensemble involves non-unitary operators that render an implementation on a near-term quantum computer extremely challenging. 
	Recently, \href{https://doi.org/10.1103/PRXQuantum.2.020321}{Lu, Ba\~nuls and Cirac, PRX Quantum 2, 020321 (2021)} suggested a ``time-series quantum Monte Carlo method'' which circumvents this problem by calculating finite temperature properties from Monte Carlo sampling of easily preparable states, where the Boltzmann weights are extracted from real-time quantum simulations via Wick's rotation. 
	In this paper, we address the challenges associated with the practical applications of this method, using the two-dimensional transverse field Ising model as a testbed.
 We demonstrate that estimating Boltzmann weights via Wick's rotation is very sensitive to time-domain truncation and statistical shot noise. To alleviate this problem, we introduce a technique that imposes constraints on the density of states, most notably its non-negativity, and show that this way, we can reliably extract Boltzmann weights from noisy time series. In addition, we show how to  reduce the statistical errors of Monte Carlo sampling via a reweighted version of the Wolff cluster algorithm. Our work enables the implementation of the time-series algorithm on present-day quantum computers to study finite temperature properties of many-body quantum systems.
	\end{abstract}

\maketitle

\begin{figure}[h]
	\center
	\includegraphics[width=0.47\textwidth]{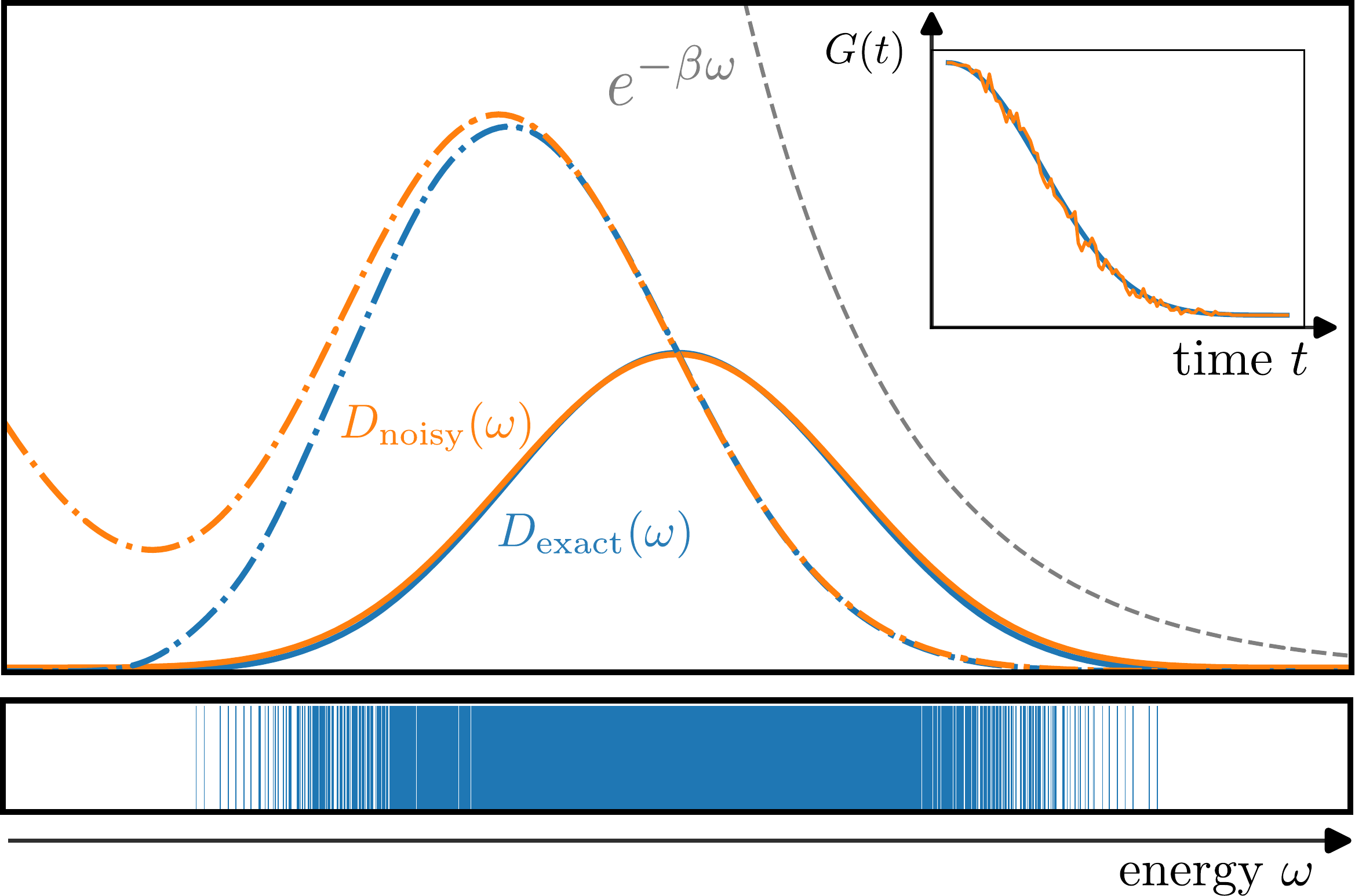}
	\caption{\label{fig:conceptual}
		Equilibrium properties from quantum dynamics and the noise-induced infrared catastrophe.
		Bottom panel: Energy spectrum of a 12-qubit Ising chain in a transverse field $h_x=0.5$. In principle, the spectrum contains sufficient information to compute properties like thermal phase transitions.
		Top panel inset: A quantum computer or simulator can be used to efficiently compute the Loschmidt amplitude $G(t) = \braket{\psi|e^{-iHt}|\psi}$ which contains information about the local density of states via its Fourier transform. In practice, determining $G(t)$ suffers from shot noise (orange).
		Top panel: Collecting data from suitably chosen initial states yields the density of states $D(w)$. While it is easy to obtain a reasonable estimate for said density (solid lines), one needs to weight $D(w)$ with a Boltzmann factor (gray dashed) to obtain thermal observables. Any small error induced by shot noise is amplified exponentially (dash-dotted orange vs. blue), leading to significant  artefacts when simulating low-energy physics.
		The present paper introduces a method to address this problem.
	}
\end{figure}

Calculating finite temperature observables of quantum systems lies at the heart of predicting material properties as they determine many experimentally relevant quantities such as the specific heat or susceptibilities. Moreover, finite temperature phase transitions separate phases of matter with radically different properties, with paradigmatic examples including the solid-liquid, ferromagnet-paramagnet and insulator-superconductor transitions. The prediction of phase transitions requires the solution of a many-body quantum problem, which in general can not be done analytically.

Several classical algorithms exist for numerically evaluating thermal observables, including the stochastic series expansion method~\cite{Sandvik91, Sandvik92},
world line quantum Monte Carlo~\cite{Suzuki77, Hirsch82}, 
density matrix quantum Monte Carlo~\cite{Blunt14, Petras20},
minimally entangled typical thermal states (METTS)~\cite{White09, Stoudenmire10},
determinantal quantum Monte Carlo~\cite{Blankenbecler81, White89},
finite temperature auxiliary field quantum Monte Carlo~\cite{Liu18, He19},
and finite temperature density matrix embedding theory~\cite{Sun20}. Despite half a century of progress on these computational methods, many interesting regimes are still inaccessible. Most notably, Monte Carlo methods struggle with frustrated spin models and fermionic systems due to the sign problem~\cite{Henelius2000, Troyer05}.

To overcome such challenges, algorithms for quantum computers have been proposed, such as the quantum Metropolis algorithm~\cite{Temme11, Yung12}, quantum imaginary time evolution~\cite{Motta20}, minimal effective Gibbs ansatz~\cite{Cohn20}, thermal pure quantum states~\cite{Sugiura12, Coopmans23} and a quantum version of METTS~\cite{Sun23}. While fault-tolerant quantum algorithms have proven improved asymptotic scalings~\cite{Yung12, Poulin09, Chowdhury17, Bravyi21, Coopmans23}, they have punishingly large resource estimates, so their implementation on current noisy intermediate scale quantum (NISQ) devices has been confined to small systems~\cite{Rusland2022, Huggins2022}.

Recently, Ref.~\cite{Lu2021} proposed a hybrid quantum-classical Markov chain algorithm for calculating thermodynamic properties, which we call \emph{time-series quantum Monte Carlo}.
The algorithm can be adapted for calculating observables in either the microcanonical or the canonical ensembles, and it inspired new classical methods that were further studied in Ref.~\cite{Yang2022} (microcanonical) and Ref.~\cite{Schuckert2022} (canonical).
In this algorithm, the Boltzmann weights, i.e., the probability of sampled states in the thermodynamic ensemble, are extracted from their real-time dynamics evaluated on the quantum computer, and thus no sign problem appears.
Ref.~\cite{Yang2022, Schuckert2022} suggested that the algorithm works well for large non-integrable spin systems by showing promising results for long chains of the transverse-field Ising model.
However,  Ref.~\cite{Schuckert2022} also pointed out that any small errors, either experimental or due to numerical imprecision, might spoil the implementation of the algorithm at finite temperatures. Moreover, the required number of Monte Carlo iterations, even on a relatively simple spin model, is very large, meaning the clock time on the quantum machine is long. 

Here, we overcome these limitations of the time-series algorithm by introducing a noise-robust method based on non-negative least squares (NNLS)~\cite{Lawson95} to extract the Boltzmann weights from the time series. Moreover, we introduce a cluster update that reduces the number of required Monte Carlo iterations by orders of magnitude. 
The structure of the paper is as follows. We first briefly introduce the time-series algorithm and the transverse field Ising model, which we use to benchmark our methods. We then discuss evaluating the Loschmidt echo, the quantity required for the time-series algorithm via quantum circuits, and how to efficiently sample product states in the transverse field Ising model using Wolff's cluster update algorithm~\cite{Wolff89}.
Next, we systematically investigate how to estimate Boltzmann weights reliably from Loschmidt echos and study the effect of various parameters on the error in estimated weights.
Finally, we perform simulations of the complete algorithm at different temperatures in the presence of shot noise using IBM's AerSimulator and discuss its results.

\section{Thermal equilibrium observables from the time-series algorithm}
All information about a system in thermodynamic equilibrium can be obtained from the expectation value
\begin{equation}
	\braket{\hat{O}} = \operatorname{Tr}\left[\hat{\rho}\ \hat{O}\right],
\end{equation} 
where $\hat{O}$ is the operator representing the observable of interest and $\hat{\rho}$ is the density operator describing the statistical ensemble with the specified thermodynamic constraints. 
In particular, we are interested in the canonical ensemble, where the temperature of the system $1/\beta$ is fixed. This is described by the density operator
\begin{equation}
	\hat{\rho} = \frac{e^{-\beta \hat{H}}}{\operatorname{Tr}\left[e^{-\beta \hat{H}}\right]}.
\end{equation}
In the time-series algorithm~\cite{Lu2021, Schuckert2022, Yang2022}, thermal expectation values are evaluated using a complete orthonormal basis of states $\ket{\psi}$
\begin{align}\label{eq:o_sum}
\braket{\hat{O}} &=  \frac{\sum_{\psi} \braket{\psi | e^{-\beta \hat{H}}\ \hat{O}| \psi}}{\sum_{\psi} \braket{\psi | e^{-\beta \hat{H}}| \psi}} \\&= \frac{\sum_{\psi} W_\psi O_\psi}{\sum_{\psi} W_\psi},
\end{align}
where the Boltzmann weights $W_\psi$ and local observables $O_\psi$ are defined as
\begin{align}
W_\psi(\beta) &\coloneqq \braket{\psi | e^{-\beta \hat{H}}| \psi} \\
O_\psi(\beta) &\coloneqq \frac{\braket{\psi | e^{-\beta \hat{H}}\ \hat{O}| \psi}}{\braket{\psi | e^{-\beta \hat{H}}| \psi}}.\label{eq:Opsi}
\end{align}
Because $e^{-\beta \hat H}$ is positive definite for any Hamiltonian $\hat H$, the Boltzmann weights are non-negative numbers that can be normalized and interpreted as a probability distribution
\begin{equation}
 p_\psi \coloneqq \frac{W_\psi}{\sum_{\psi^\prime} W_{\psi^\prime}}\;.
\end{equation}
The expression in Eq.~\eqref{eq:o_sum} is then estimated by sampling $M$ states from this distribution via some Monte Carlo method  and averaging over the local observable of the sampled states $\psi_i$
\begin{equation}
	\braket{\hat{O}} \approx \frac{1}{M} \sum_{i=1}^M O_{\psi_i}\;.
\end{equation}
The error of this estimator is  $ \mathcal{O}(1/\sqrt{M})$. In particular, the error only scales with the inverse square root of the number of independent samples and not the size of the Hilbert space~\cite{Foulkes01}. Nevertheless, when samples are correlated,  the correlation length can generally depend on the size of the physical system.

In ``classical'' quantum Monte Carlo methods~\cite{Gubernatis16}, the Boltzmann weights are calculated by splitting the operator $e^{-\beta \hat{H}}$ into $n$ smaller pieces $e^{-\beta \hat{H}/n}$ and inserting resolutions of the identity in between. 
The resulting terms are then calculated and summed over also by means of Monte Carlo methods.
However, unlike the original Boltzmann weights, the individual terms can generally be negative because they involve transition matrix elements between \emph{different} basis states $\braket{\psi'|e^{-\beta \hat{H}/n}|\psi}$, leading to the infamous sign problem.
To avoid this issue, the Boltzmann weights in the time-series algorithm are instead calculated by evaluating their real-time counterpart, known as Loschmidt echos, on a quantum computer
\begin{equation}
	G_\psi(t) \coloneqq \braket{\psi | e^{-iHt} | \psi}
\end{equation}
and then numerically performing a Wick's rotation to the imaginary-time axis $t \to -i\beta$.
In practice, this is done with the help of the local density of states
\begin{equation}
	D_\psi(\omega) \coloneqq \braket{\psi| \delta(\hat H-\omega) |\psi},
\end{equation}
which is nothing but the Fourier transform of Loschmidt echos
\begin{equation}\label{eq:fourier}
	D_\psi(\omega) =  \int \frac{dt}{2\pi}\ e^{i\omega t} \  G_\psi(t).
\end{equation}
Boltzmann weights are then calculated as
\begin{equation}\label{eq:integrate}
	W_\psi =\int d\omega\ e^{-\beta \omega}\ D_\psi(\omega).
\end{equation}
The last two relations can be readily verified by expressing each of the quantities in terms of the eigenstates  of the Hamiltonian $\ket{n}$ and their energies $E_n$
\begin{align}\nonumber
	D_\psi(\omega) &=   \sum_n |\braket{\psi|n}|^2\ \delta(E_n -\omega),\\\nonumber
	G_\psi(t) &= \sum_n |\braket{\psi|n}|^2 \ e^{-i E_n t},\\
	W_\psi(\beta) &= \sum_n |\braket{\psi|n}|^2 \ e^{-\beta E_n}.
\end{align}
The last equation also explicitly shows the positivity of the Boltzmann weights $W_\psi(\beta)$.
Note that as opposed to the rotation from the imaginary-time axis to the real-time axis, known as the analytic continuation problem~\cite{Silver90, Jarrell96, Gunnarsson07, Ghanem20a, Ghanem20b, Ghanem23}, obtaining the density of states from Loschmidt echos is a well-defined problem since Fourier transform is just a unitary operation. Nevertheless, obtaining Boltzmann weights from the density of states remains a difficult problem (see Fig.~\ref{fig:conceptual}), as will become evident later.
The numerator of the local observables $O_\psi$ can  be  similarly calculated with the help of a quantum computer.
In the special case where the basis $\ket{\psi}$ diagonalizes the observable $\hat{O}$, its local observables can be directly and efficiently evaluated  on a classical computer.   
In this work, we focus on the latter case.

Directly sampling the states from the Boltzmann distribution is generally not possible.
Instead, Metropolis-Hastings algorithm is a general Markov chain method which enables sampling from arbitrary distributions.
The price we pay in Markov chain methods compared to direct sampling is that these samples are correlated, and we generally need more samples to achieve the same level of accuracy.
The Metropolis-Hastings algorithm can start from any initial state, but it helps to choose a state that already has a large Boltzmann weight.
The algorithm then repeatedly proposes a new state $\ket{\psi'}$ from the current one $\ket{\psi}$ using some probability distribution $P_{\psi \to \psi'}$.
The proposed sample is accepted or rejected using an acceptance ratio 
\begin{equation}\label{eq:acceptance}
	A \coloneqq \min\left(1, \frac{W_{\psi^\prime}} {W_{\psi}} \frac{P_{\psi^\prime \to \psi}}{P_{\psi \to \psi^\prime} }\right).
\end{equation}
If accepted, the current state $\ket{\psi}$ is replaced by the new one $\ket{\psi'}$. Otherwise, the current state $\ket{\psi}$ is used.
In principle, one can use any proposal strategy as long as it guarantees ergodicity, i.e., it allows going from any state to any other state in a finite number of steps.
In practice, one should strive to use a proposal distribution that closely resembles the target distribution, here the Boltzmann distribution, in order to minimize the correlation between the samples.

\begin{figure}[t]
	\center
	\includegraphics[width=0.5\textwidth]{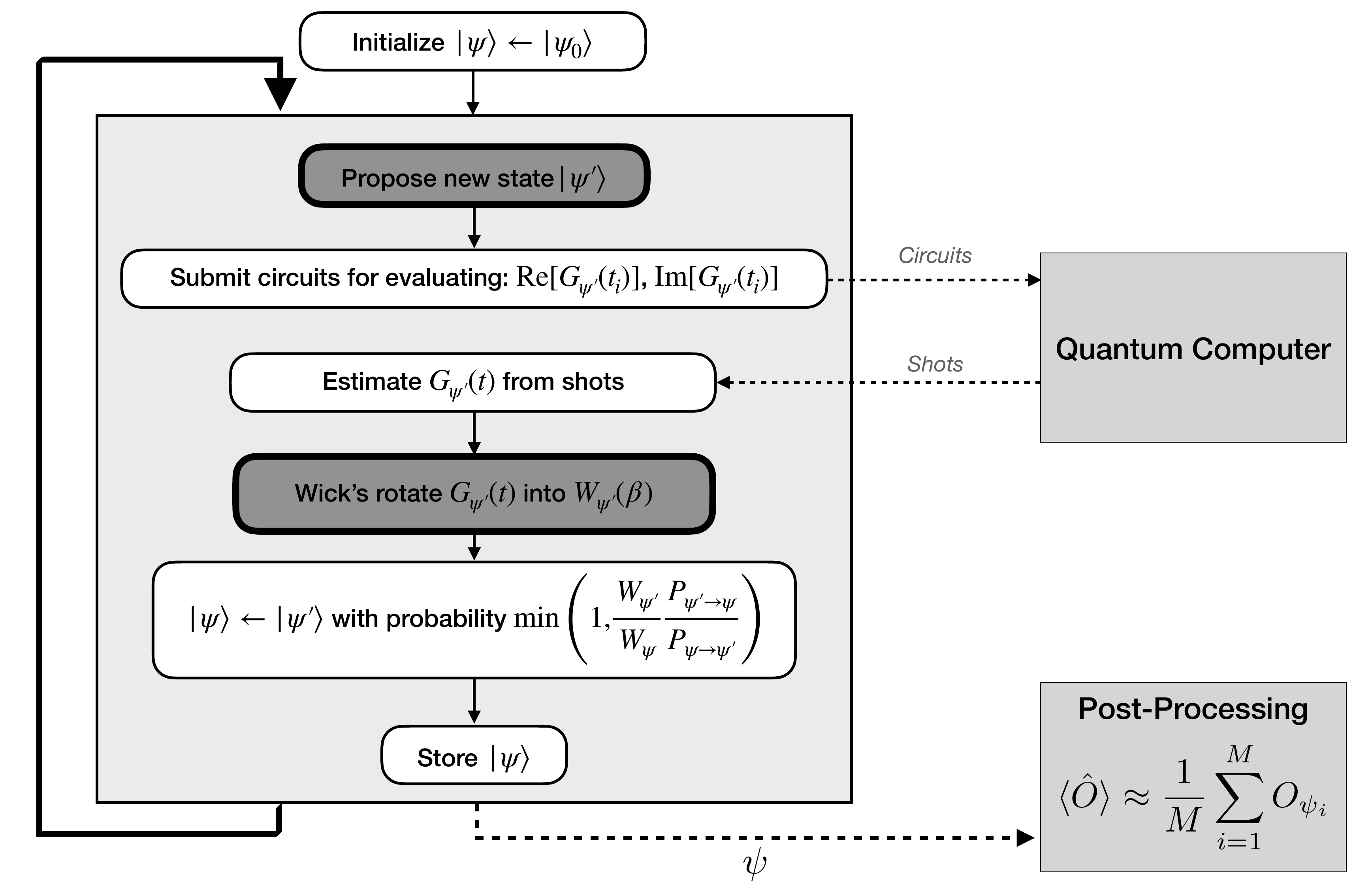}
	\caption{\label{fig:flowchart} 
		Flowchart of the time-series Monte Carlo algorithm proposed by Refs.~\cite{Lu2021} and benchmarked in Refs.~\cite{Schuckert2022, Yang2022} for calculating finite temperature properties with the help of quantum computers.
        In this work, we provide improvements on the highlighted steps of Monte Carlo sampling and Wick's rotation. 
	}
\end{figure}

In Fig.~\ref{fig:flowchart}, we show a flow chart of the full method. The main loop of the algorithm starts by proposing a new state.
Then it submits circuits for calculating a time series of its Loschmidt echos to the quantum computer and waits for the resulting shots.
From these shots, Loschmidt echoes are estimated and used to calculate the Boltzmann weight of the new state.
We then use that to accept or reject the proposed state.
Finally, at the end of an iteration, we store the state for later post-processing and calculations of observables.
In the rest of this paper, we will discuss each of these steps in detail, taking the transverse-field Ising model as a specific example.

\section{Test Case: Transverse-Field Ising Model}
\begin{figure}[t]
	\center
	\includegraphics[width=0.9\columnwidth]{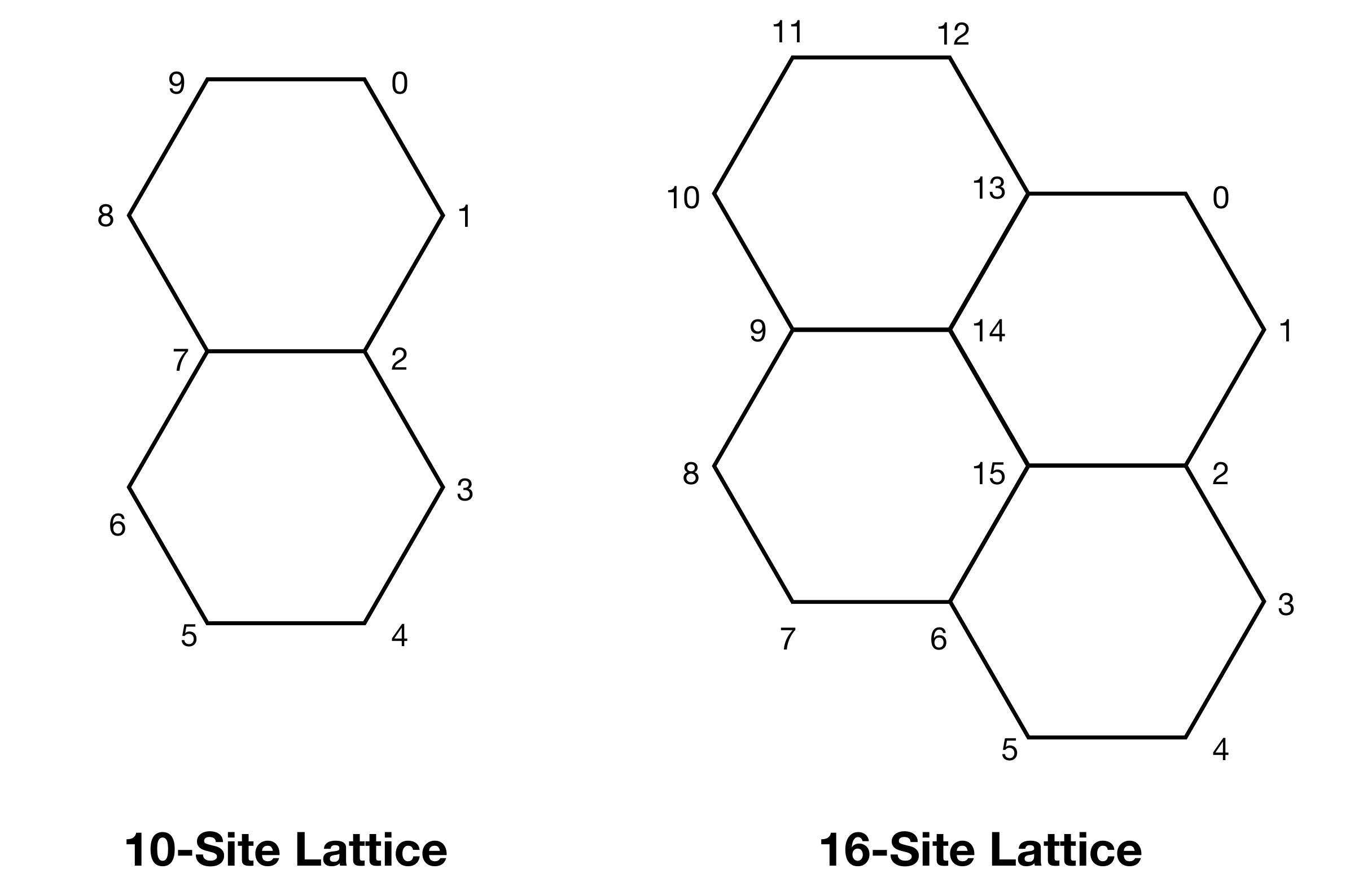}
	\caption{\label{fig:lattices} 
		The two honeycomb lattices explored in this work.
		The numbers represent the indices of the sites.
		When implementing the time evolution quantum circuits, the depth is reduced by executing the gates corresponding to the bonds in each of the three different orientations in parallel.
		For example, in the 10-site lattice, the following sets of bonds can parallelized: $\{(0, 1), (2, 3), (5, 6), (7, 8)\}$, $\{(1, 2), (3, 4), (6, 7), (8, 9)\}$ and $\{(0, 9), (2, 7), (4, 5)\}$.
	}
\end{figure}

The transverse-field Ising model (TFIM) consists of a set of quantum magnetic dipoles of spin 1/2 on a lattice with nearest neighbor interactions. 
The interaction is determined by the spin alignment along some axis (usually taken to be the z-axis) in addition to an external magnetic field along a perpendicular axis (usually the x-axis). 
Its Hamiltonian on  a general lattice reads
\begin{equation}
	{\displaystyle \hat{H} =-J \sum_{\langle i,j\rangle }\hat{Z}_{i}\hat{Z}_{j}+ h_x\sum _{i}\hat{X}_{i}}\;,
\end{equation} 
where $\langle i,j\rangle $ indicates summation over neighboring lattice sites, $J$ is the coupling constant, which we set  in the following to $J=1$, specifying ferromagnetic interactions. $h_x$ is the strength of the external filed, and $\hat{Z}_i$ and $\hat{X}_i$ are Pauli matrices representing, respectively, the z- and x-components of the spin-$\frac{1}{2}$ operator at lattice site $i$.
This model is one of the simplest quantum spin models exhibiting non-trivial physics that cannot be described classically.
On the other hand, its dynamics can be relatively straightforward to implement on gate-based quantum computers, which makes it an ideal candidate for testing the algorithm.
The observable of interest is the magnetization at different temperatures $\beta$ and different values of the external field parameter $h_x$.

The TFIM can be solved exactly in 1D, where it shows a phase transition at zero temperature but no order at finite temperatures~\cite{Pfeuty70}.
In higher dimensions, the phase diagram is divided by a second-order critical line into a ferromagnetic ordered region with $\braket{\hat{Z}} \neq 0$ and paramagnetic disordered one with $\braket{\hat{Z}} = 0$~\cite{Friedman78}.
The endpoints of the critical line are characterized by the critical temperature $\beta_c$ of the classical Ising model (i.e., $h_x=0$) and a critical value of the field parameter  at zero temperature.
In this work, we use a two-dimensional honeycomb lattice with 10 and 16 sites (see Fig.~\ref{fig:lattices}).
The classical critical temperature of this lattice is $ \beta_c = \frac{\log{(2+\sqrt{3})}}{2}$ in the thermodynamic limit~\cite{Wannier45, Morita16}.
As a sampling basis set $\ket{\psi}$, we choose the product states of the $Z_i$ operators.
These are the easiest states to prepare on a quantum computer and can be directly implemented using $X$ gates.
Besides, they are the eigenstates of the magnetization operator; therefore, we can easily calculate their local magnetization on a classical computer and only need the quantum computer for evaluating Loschmidt echos.

\section{Quantum Circuits}
\begin{figure}[t]
	\center
	\includegraphics[width=0.84\columnwidth]{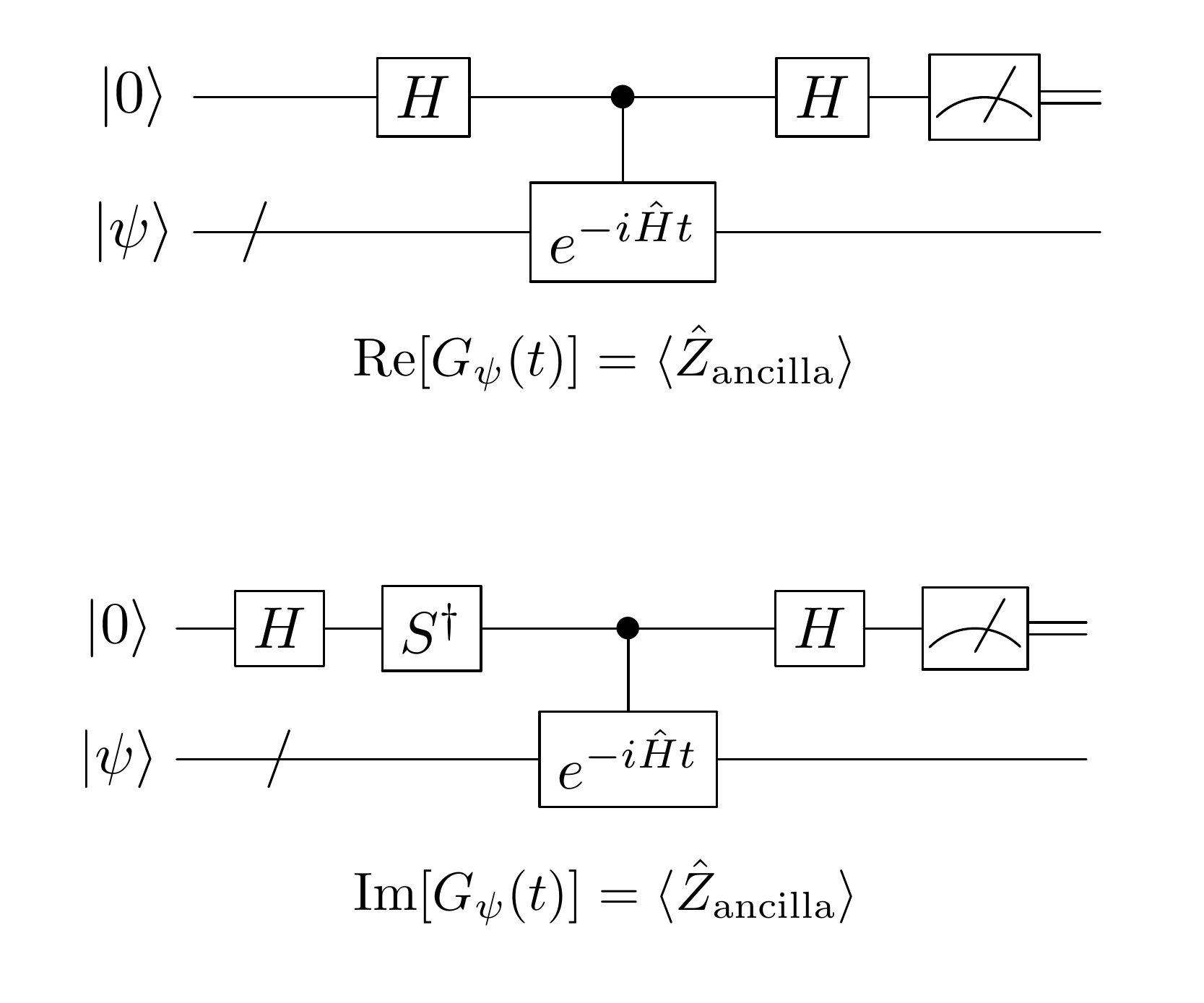}
	\caption{\label{fig:hadamard_test} 
		Hadamard test circuits for calculating the real (top) and imaginary (bottom) parts of Loschmidt echo.
		The box $e^{-i\hat{H} t}$ represents the time evolution circuit (see Fig.~\ref{fig:time_evolution}), $H$ is the Hadamard gate and $S^\dagger$ is the adjoint of the phase gate.
	}
\end{figure}

Calculating the real and imaginary parts of Loschmidt echos can be achieved using the so-called Hadamard test.
This involves two quantum circuits, one for the real part and one for the imaginary part, where each circuit contains one ancilla qubit and a time evolution circuit controlled by the ancilla as shown in Fig~\ref{fig:hadamard_test}. 
By measuring the ancilla qubit, the desired real and imaginary parts can be calculated as the expectation value of the observable $\hat{Z}_\text{ancilla}$, i.e., the difference between the probability of measuring the value 0 and the probability of measuring the value 1. Moreover, by additionally measuring $ \hat Z_\text{ancilla} \otimes \hat O $, i.e. correlating a measurement of $\hat O$ on the system qubits with a measurement of $\hat Z$ on the ancilla, we can obtain $\braket{\psi|\hat O e^{-i\hat H t}|\psi}$ additionally to the Loschmidt echos from the same shots~\cite{Lu2021}. This enables the evaluation of Eq.~\eqref{eq:Opsi} for arbitrary observables. As mentioned before, here we focus on the case where this is not necessary, since $\ket{\psi}$ are eigenstates of $\hat O$.

The time evolution circuits are obtained via second-order Trotterization.
\begin{equation}\label{eq:trotter}
	e^{-i\hat{H} t} \approx \left[e^{-i\hat{A} t/2n}  e^{-i\hat{B} t/n}  e^{-i \hat{A} t/2n} \right]^{n}.
\end{equation}
where the Hamiltonian is split into two non-commuting parts $\hat{H} = \hat{A}+\hat{B}$. The first part $\hat{A}$ contains the $\hat{X}_i$ terms, while the second $\hat{B}$ contains the mutually commuting $\hat{Z}_i \hat{Z}_j$ terms.
The time evolution of each part can be achieved easily using one- and two-qubit gates.
In particular, we add one $R_{ZZ}(-2J t)$ for each pair of interacting sites and an $R_X(2 ht)$ gate for each site.
To reduce the depth of the time-evolution circuits, it is important to order the $\hat{Z}_i \hat{Z}_j$ terms such that bonds with disjoint sites can be executed in parallel.
For the honeycomb lattice, this is achieved by grouping the bonds along each of the directions $0, \pi/3, 2 \pi/3$ (e.g., see the caption of Fig.~\ref{fig:lattices}).

\begin{figure}[t]
	\centering
	\subfloat[]{
		\includegraphics[width=\columnwidth]{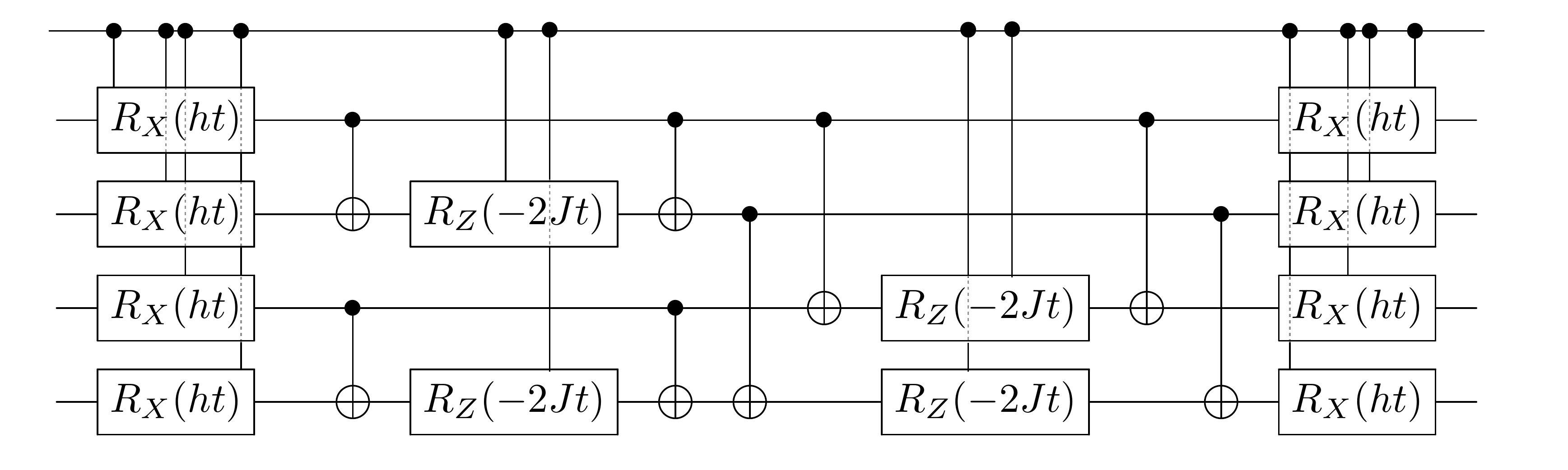}\label{fig:time_evolution} 
		
	}\\
	\subfloat[]{
		\includegraphics[width=0.6\columnwidth]{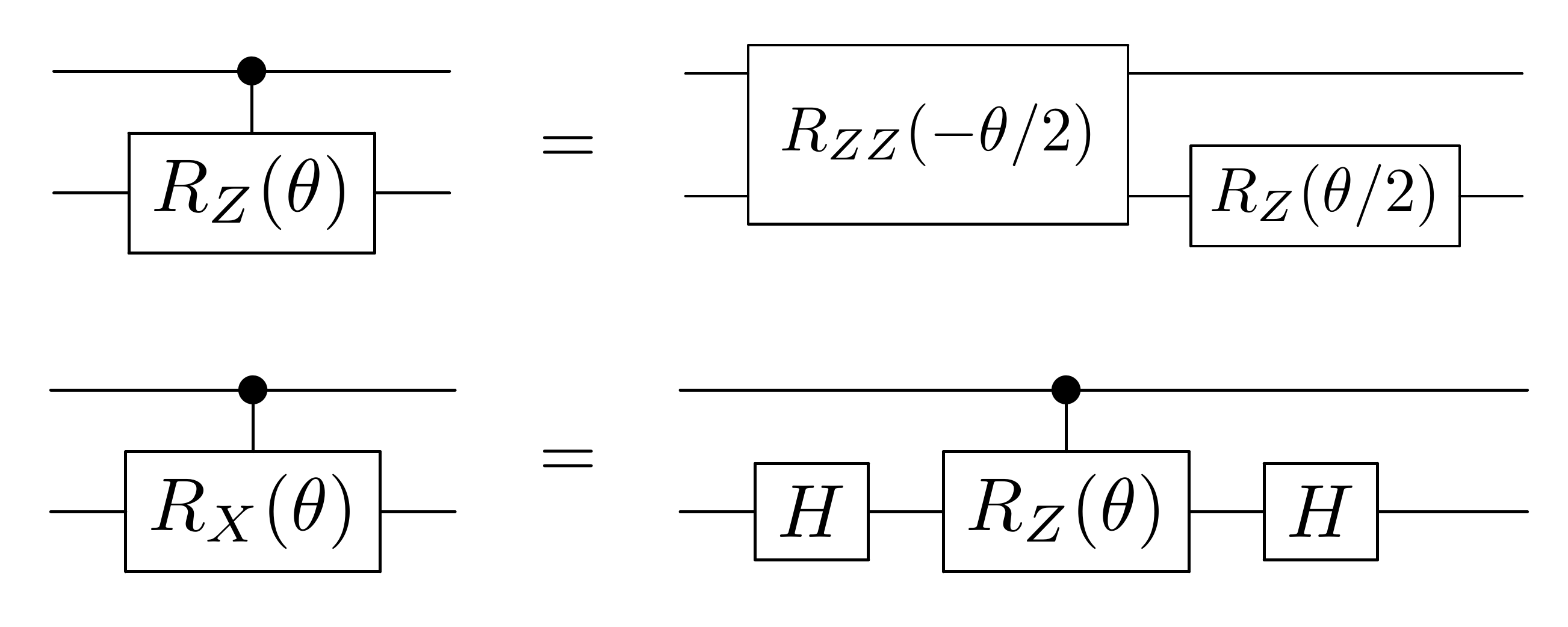}\label{fig:controlled_gates} 
	}
	\caption{
		(a) Example of the controlled time-evolution circuit for the transverse field Ising model.
		Here is shown a single second-order Trotter step for a 4-sites ring. 
		(b) Decomposition of the controlled $R_Z$ and $R_X$ gates using $R_{ZZ}$ gate and other single-qubit gates.
	}
\end{figure}

The Hadamard test requires controlling the time evolution circuit, which is achieved by controlling the $R_{ZZ}$ and $R_X$ gates.
To control an $R_{ZZ}$ gate, we first decompose it into an $R_Z$ gate sandwiched between two CNOT gates and then control only the $R_Z$ gate.
CNOT gates do not need to be controlled because they always come in pairs.
The resulting controlled $R_X$ and $R_Z$ gates are implemented each using a single $R_{ZZ}$ gate, which is native to ion-trapped quantum computers, as shown in Fig~\ref{fig:controlled_gates}.
Consequently,  a single time slice of $\hat{A}$ requires a number of two-qubit gates that equals the total number of lattice sites, while implementing a single slice of $\hat{B}$ requires three times the number of bonds in the lattice.
An example of the controlled time-evolution circuit is given in Fig.~\ref{fig:time_evolution}.

To reduce the number of two-qubit gates in the controlled circuit, we rewrite the second-order Trotterization as following
\begin{equation}
	e^{-i\hat{H}t} \approx e^{-i\hat{A} t/2n}  \left[e^{-i\hat{B} t/n} e^{-i\hat{A} t/n}  \right]^{n}  e^{i\hat{A} t/2n} .
\end{equation}
By implementing $e^{-i\hat{A} t/2n}$ as part of the state preparation circuit, $\hat{A}$ terms appear only $n$ times in the controlled time evolution circuit instead of the $n+1$ times needed using  Eq.~\eqref{eq:trotter} .
In total, implementing a single controlled second-order  Trotter step requires 43 two-qubit gates for the 10-site lattice and 73 for the 16-site one.

\section{Monte Carlo Sampling}\label{sec:monte_carlo}
The most straightforward strategy for sampling product states of the TFIM is the single spin flip update.
This strategy picks a site randomly from the lattice and reverses its spin.
The proposal probability is symmetric and equals $P_{\psi^\prime \to \psi} = P_{\psi \to \psi^\prime} = 1/L$.
Therefore, the acceptance ratio is determined solely by the Boltzmann weights:
\begin{equation}
	A = \min\left(1, \frac{W_{\psi^\prime}} {W_{\psi}} \frac{P_{\psi^\prime \to \psi}}{P_{\psi \to \psi^\prime} }\right) = \min\left(1, \frac{W_{\psi^\prime}} {W_{\psi}}\right).
\end{equation}
However, the old and new states differ only by one spin, which means that this local strategy is slow in exploring the Hilbert space, leading to a high correlation at low and moderate temperatures.

\begin{figure}[t]
	\center
	\includegraphics[width=\columnwidth]{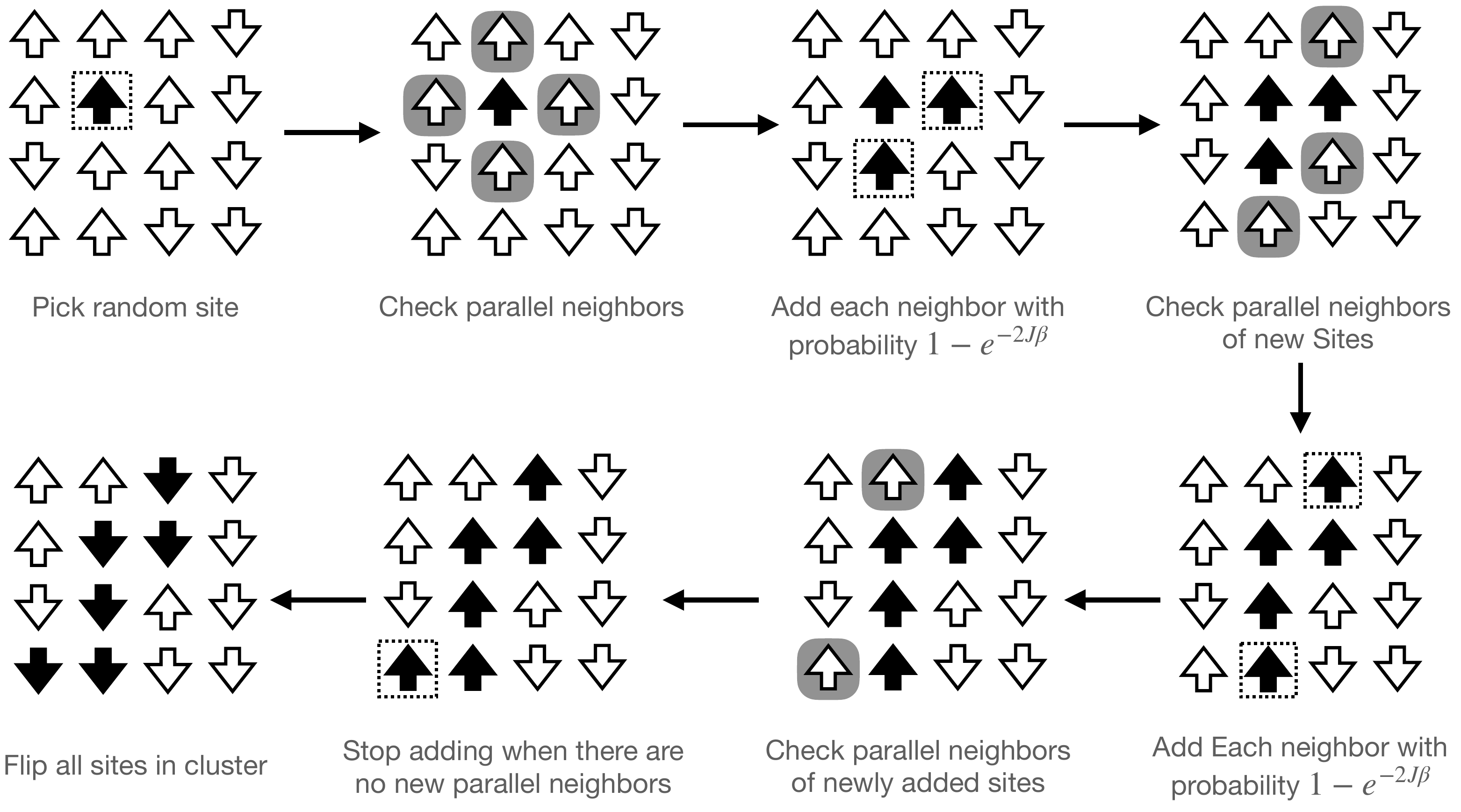}
	\caption{\label{fig:cluster_update} 
		Example of a cluster update step taken by Wolff's algorithm. 
		Arrows represent sites of spin up or down, and black ones denote those already added to the cluster.
		Dashed boxes designate the most recently added sites, while gray boxes designate new neighbors with the same spin as the cluster.
	}
\end{figure}

A more effective approach that makes global moves is Wolff's cluster update algorithm~\cite{Wolff89}.
It involves randomly selecting a cluster of aligned spins and flipping all of the spins in the cluster at once.
The cluster is constructed by choosing a random lattice site (the cluster's root) and adding with probability $1-e^{\beta J}$ each of the neighboring spins that point in the same direction.
The parallel neighbors of the newly added sites are again added with that probability, and the cluster keeps expanding stochastically until there are no new parallel neighbors. An example on a square lattice is given in Fig.~\ref{fig:cluster_update}.

This algorithm is designed to sample from the Boltzmann distribution of the classical Ising model, and it was shown to be efficient even near the critical temperature~\cite{Wolff89}.
Here, we adapt this algorithm by using it as a proposal strategy within the Metropolis-Hastings algorithm.
To this end, we need the ratio of the proposal probabilities, which is equal to the ratio of the classical Boltzmann weights of the new and old states~(see Eq.~(6) of Ref.~\cite{Wolff89}). Given that the energies of the product states in the classical Ising model are the same as those of TFIM, this ratio can be written as
\begin{equation}
	\frac{P_{\psi^\prime \to \psi}}{P_{\psi \to \psi^\prime} } = \frac{e^{-\beta \braket{\psi| \hat{H} | \psi}}}{e^{-\beta \braket{\psi^\prime| \hat{H} | \psi^\prime}}},
\end{equation}
which involves only energies of the product states and can be calculated efficiently on the classical computer.
We want to emphasize here that using the cluster update as a proposal strategy does not affect the correctness of the Monte Carlo simulations.
As a direct application of the Metropolis-Hastings algorithm, these simulations satisfy the detailed balance condition for the quantum TFIM. In the Metropolis-Hastings algorithm, the detailed balance condition is satisfied as long as the proposed samples are accepted with the proper acceptance ratio (Eq.~\eqref{eq:acceptance}).
One is then free to choose any ergodic proposal strategy as long as the correct
ratio of proposal probabilities is used in the acceptance ratio. 
When using the cluster update, the acceptance ratio reads
\begin{equation}
	A = \min\left(1, \frac{W_{\psi^\prime}} {W_{\psi}} \frac{e^{-\beta \braket{\psi| \hat{H} | \psi}}}{e^{-\beta \braket{\psi^\prime| \hat{H} | \psi^\prime}}} \right).
\end{equation}

\begin{figure}[t]
	\center
	\includegraphics[width=\columnwidth]{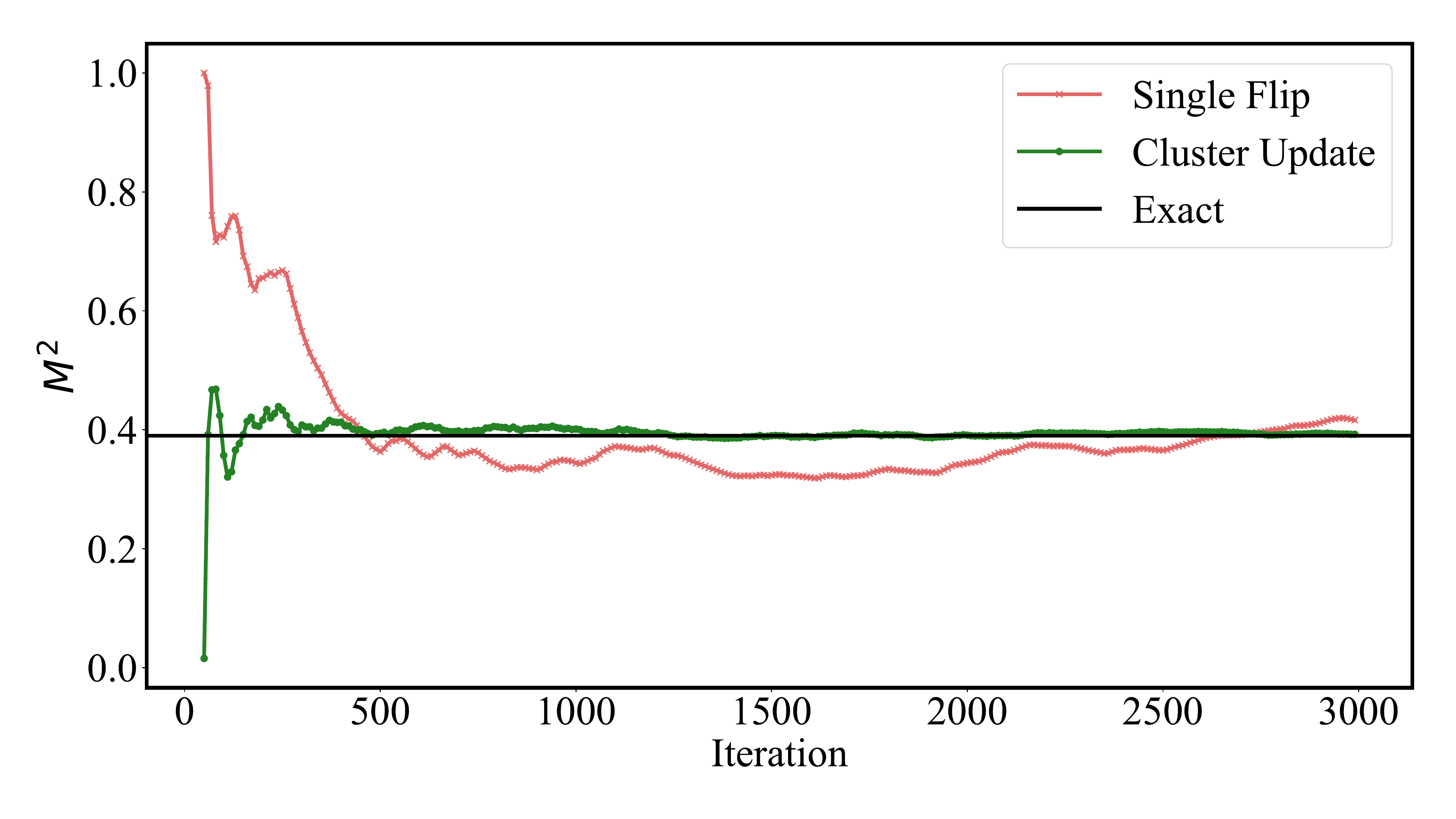}
	\caption{\label{fig:sampling_comparison} 
		Evolution of the average squared magnetization using different proposal strategies.
		Each point represents the average of the chain up to the specified iteration.
		The first $50$ iterations (burn-in period) are excluded.
		Both chains start with the all-spin-down state.
	}
\end{figure}

In Fig.~\ref{fig:sampling_comparison}, we compare the single flip update with the cluster update for the 16-site honeycomb lattice with $h_x=0.5$ at the classical critical temperature $\beta_c$.
The plot demonstrates how the cluster update thermalizes almost instantly, while the single flip update needs hundreds of iterations and suffers from much longer correlation times.
The efficiency of the cluster update depends on the strength of the field $h_x$ and inverse temperature $\beta$, because the classical Boltzmann distribution, which is used to propose samples in the cluster update, deviates more from the target quantum Boltzmann distribution at lower temperatures and higher $h_x$.
This can be seen by expanding classical and quantum Boltzmann weights as a function of inverse temperature
\begin{equation}
    \begin{split}
    e^{-\beta \braket{\psi| \hat{H}|\psi}} &= 1- \beta \braket{\psi| \hat{H} | \psi} + \sum_{n=2}^\infty \frac{(-\beta)^n}{n!} \braket{\psi| \hat{H} | \psi}^n\;, \\
    \braket{\psi|e^{-\beta \hat{H}}|\psi} &= 1 - \beta \braket{\psi| \hat{H} | \psi} + \sum_{n=2}^\infty \frac{(-\beta)^n}{n!} \braket{\psi| \hat{H}^n | \psi}\;.
    \end{split}
\end{equation}
When $h_x\neq 0$, the classical and quantum moments of product states do not match $\braket{\psi| \hat{H} | \psi}^n \neq \braket{\psi| \hat{H}^n | \psi}$ for $n\geq 2$, and the difference is proportional to $h_x^n$. Consequently, sampling with the cluster update becomes less efficient for lower temperatures and higher $h_x$. 

\section{Wick's Rotation}
Wick's rotation is the most crucial component of the time-series Monte Carlo algorithm.
The procedure for performing it is formally straightforward: Compute the local density of states $D_\psi(\omega)$ via Fourier transform of Loschmidt echos [cf. Eq.~\eqref{eq:fourier}], then integrate it with $e^{-\beta \omega}$ factor to get Boltzmann weights [cf. Eq.~\eqref{eq:integrate}].
In practice, however, obtaining Boltzmann weights reliably is difficult, especially for lower temperatures.
The problem comes from the exponential factor $e^{-\beta \omega}$, which makes Boltzmann weights sensitive to the errors made in estimating the local density of states.
Therefore, we need to find a way to obtain the Fourier transform with a high accuracy using only a limited number of noisy data points.

In this section, we first discuss the issue of only having a limited number of data points using the method of Gaussian filtering introduced by Ref.~\cite{Schuckert2022}.
We propose an automatic way of determining its parameters and study its error behavior.
We then introduce the non-negative least squares (NNLS) method, which gives superior results, especially for a small number of data points.
Next, we address the issue of statistical shot noise which is present in any quantum computer (even a fault-tolerant one) and show how to stabilize NNLS in the presence of this noise.
Finally, we benchmark our recipe using the 10-site TFIM.

\subsection{Effect of Time-Series Truncation}

Given that the spectrum is bounded, then according to the Nyquist–Shannon  theorem~\cite{Shannon49}, we only need to calculate Loschmidt echos at discrete times spaced at an interval 
\begin{equation}\label{eq:nyquist}
	\Delta t \leq \frac{\pi}{\text{max} \left\{ |E_i| \right\}}.
\end{equation}
The spectral bound $\text{max} \left\{ |E_i| \right\}$ depends on the system under study. For example, the 10- and 16-sites TFIM  with $h_x=1$ has the following values $13.4478$ and $22.6223$, respectively.
Obtaining the exact value of this bound requires solving for the lowest and highest eigenenergies of the Hamiltonian $\hat{H}$, which is a difficult problem in general. However, in many cases, one can already obtain an upper bound analytically. For example, for the TFIM, we find the following upper bound
\begin{equation}
\begin{split}
\text{max} & \left\{ |E_i| \right\} = \max_{|\psi|^2 = 1} \ |\braket{\psi|\hat{H}|\psi}|  \\ 
&\leq \max_{|\psi|^2 = 1} \ |J| \sum_{\langle i,j\rangle } \braket{\psi|\hat{Z}_{i}\hat{Z}_{j}|\psi}+ |h_x| \sum _{i} |\braket{\psi|\hat{X}_{i}|\psi}| \\
&\leq |J| N_\text{bonds} + |h_x| N_\text{sites}
\end{split}
\end{equation}
where $N_\text{bonds}$ is the number of lattice bonds and $N_\text{sites}$ is the number of lattice sites.

\begin{figure}[t]
	\includegraphics[width=\columnwidth]{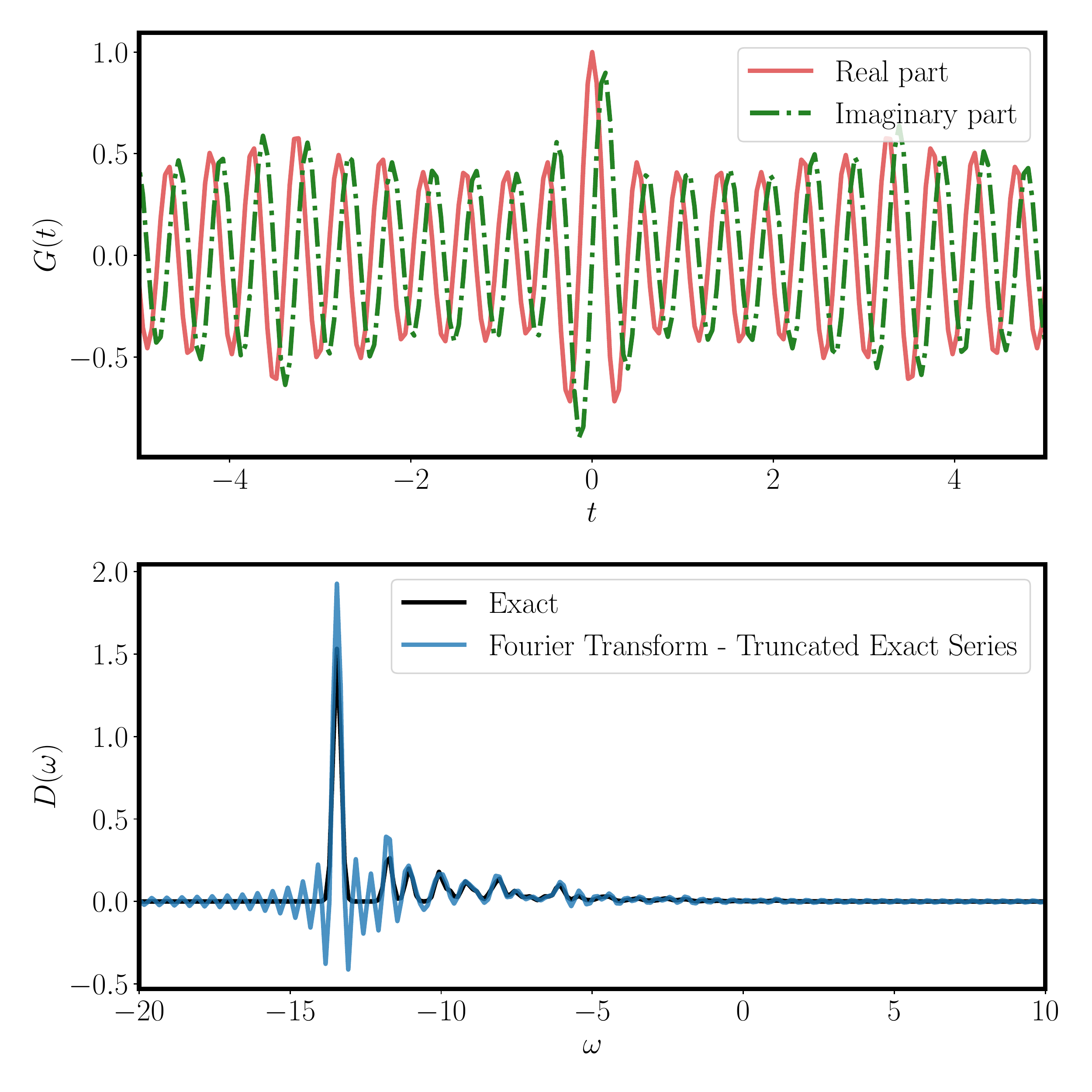}
	\caption{\label{fig:spectra} 
		Illustration of the effect of truncating the time series on its Fourier transform.
		The top panel shows the time series of the all-spin-up state of 10-sites TFIM with $h_x=1$.
		The bottom panel shows in blue its Fourier transform when sampled at a rate $1/\Delta t = 64/\pi$ and truncated at $T_\text{max} = 4 \pi$.		
		The exact density of states, a linear combination of delta functions located at the eigenenergies, is plotted in black.
		For ease of visualization, the delta functions are replaced by Gaussians whose width equals the grid spacing $\Delta \omega = 1/8$. Note the negative component of the density of states obtained from the truncated time series.
	}
\end{figure}

Although it is sufficient to sample Loschmidt echo at the finite rate of Eq.~\eqref{eq:nyquist}, we still need, in principle, to evaluate them at an infinite number of times points. Otherwise, truncating the time series at a maximum time $T_\text{max}$ is equivalent to convoluting its Fourier transform with a sinc function
\begin{equation}
	D_{\psi}^{T_\text{max}}(\omega) =   \int d\omega^\prime\ D_\psi(\omega-\omega^\prime)\ \frac{\sin(T_\text{max}\ \omega^\prime)}{ \pi \omega^\prime}.
\end{equation}
An illustrative example is shown in Fig.~\ref{fig:spectra}.
This convolution is problematic for three reasons. On the one hand, the new truncated density $D_{\psi}^{T_\text{max}}$ is not band limited, so we need to choose $\delta t$ smaller in order to evaluate its values at higher $|\omega|$.
On the other hand, even with a full knowledge of $D_{\psi}^{T_\text{max}}$, the integral of the Boltzmann weights will not converge because $1/\omega$ decays much slower than the exponential increase of $e^{-\beta \omega}$ for large negative frequencies.
Finally, even for integrals that do converge analytically, e.g., when $\beta = 0$ or for the microcanonical algorithm~\cite{Lu2021}, the oscillations of the sin function make detecting the convergence numerically more challenging.

\subsection{Gaussian Filter Method}
To circumvent these issues, Ref.~\cite{Schuckert2022} proposed to multiply the time series by an (unnormalized) Gaussian function of width $1/\delta \ll T_\text{max}$ 
\begin{equation}
	G_{\psi, \delta}(t) = G_\psi(t)\ e^{-\delta^2 t^2/2}.
\end{equation}
This is equivalent to convolving its Fourier transform with a Gaussian of the reciprocal width $\delta$
\begin{equation}
	D_{\psi, \delta}(\omega) = \frac{1}{\delta \sqrt{2\pi}}\int d\omega^\prime\ D_\psi(\omega-\omega^\prime)\ e^{- \frac{{\omega^\prime}^2}{2 \delta^2}}.
\end{equation}
Inserting this ansatz into Eq.~\eqref{eq:integrate}, we find that the effect of this broadening on the Boltzmann weights is simply a multiplication by a scale factor
\begin{equation}
	W_{\psi, \delta}(\beta) \coloneqq \int d\omega\ e^{-\beta \omega} D_{\psi, \delta}(\omega) =  e^{\beta^2 \delta^2/2}\ W_\psi(\beta). 
\end{equation}
Since this scale factor is independent of the specific state $\ket{\psi}$, the normalized Boltzmann weights $p_\psi$ stay invariant.
While the broadened density in its exact form produces an equivalent result, it is more amenable to approximations.
For large enough $\delta \gg 1/T_\text{max}$, the Gaussian filter smoothens out the oscillations in the region of interest in $D_{\psi,\delta}(\omega)$, which allows cutting off the remaining artificial oscillations in the tail.
Ref.~\cite{Schuckert2022} proposes setting to zero all density values below a certain positive cutting threshold $D_{\psi}^\text{cut}$.
However, it does not discuss how to choose such a threshold or what error the truncation introduces in the resulting Boltzmann weights.

Ideally, we want to choose the minimum cut such that it filters out the artificial oscillations but nothing more.  
Determining this value exactly is hard because it requires  disentangling the original peaks from the artificial ones.
Instead, we propose, as a heuristic, to use the maximum absolute value of the negative part of the density to determine the magnitude of those oscillations.
We then choose the cut proportional to this value
\begin{equation}\label{eq:cut}
	D_{\psi}^\text{cut} = C^\text{cut} \times \max \{|D_{\psi}(\omega)| : D_{\psi}(\omega)  < 0\}\;.
\end{equation}
We found that using $C^\text{cut}=2$ gives stable results in all examined cases.

\begin{figure}[t]
	\center
	\includegraphics[width=\columnwidth]{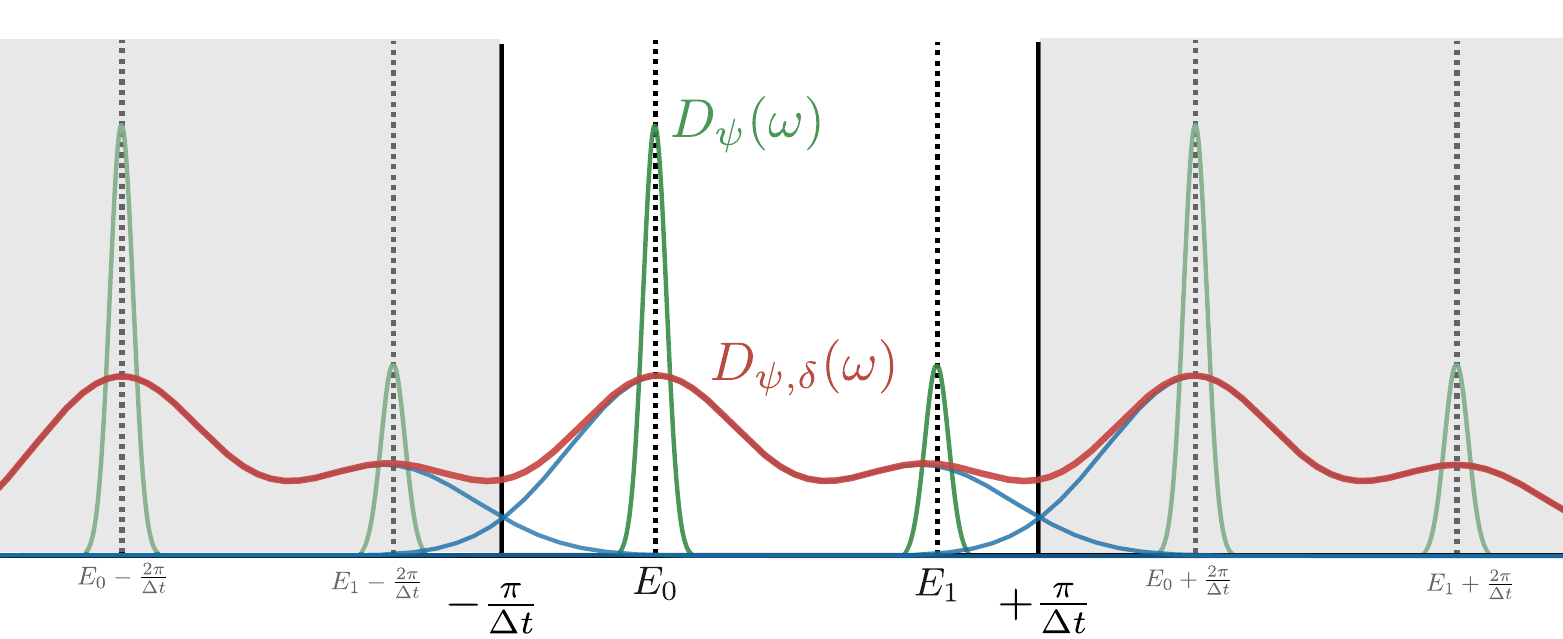}
	\caption{\label{fig:interference} 
        Schematic plot of premature truncation and interference effects.
        Using a finite sampling frequency $\pi/\Delta t$, the Fourier transform is a periodic summation of the original density (green) with period $2\pi/\Delta T$.
        As $\delta$ increases, the broadened Fourier transform (red) inside the window $[-\pi/\Delta t, \pi/\Delta t[$ increasingly deviates from the true broadened density (blue) due to interference with shifted copies outside this window.
        The finite sampling frequency  $\pi/\Delta T$, thus, puts a practical limit on the width of the Gaussian filter.
	}
\end{figure}

The next question to address is choosing the parameter $\delta$ given a fixed maximum time $T_\text{max}$.
If we know the truncated time series analytically (i.e., knowing its values to infinite precision at all times prior to $T_\text{max}$), then higher values of $\delta$ are always better, and the truncation error of the Gaussian filter will decrease monotonically. 
In practice, however, there are two limitations.
The first limitation is that, unlike the original density $D_\psi$, the broadened one $D_{\psi, \delta}$ is not band limited but decays as a Gaussian.
Therefore, one must choose a large enough $\omega$ window outside which the weight is negligible.
This window increases as $\delta$ increases, and thus to maintain the same level of accuracy, we need to decrease the time step $\Delta t$.
Failing to do so will lead to premature truncation of the density and an interference with shifted copies of itself (see Fig.~\ref{fig:interference}).

\begin{figure}[t]
	\centering
	\subfloat[Double Precision]
	{
		\includegraphics[width=\columnwidth]{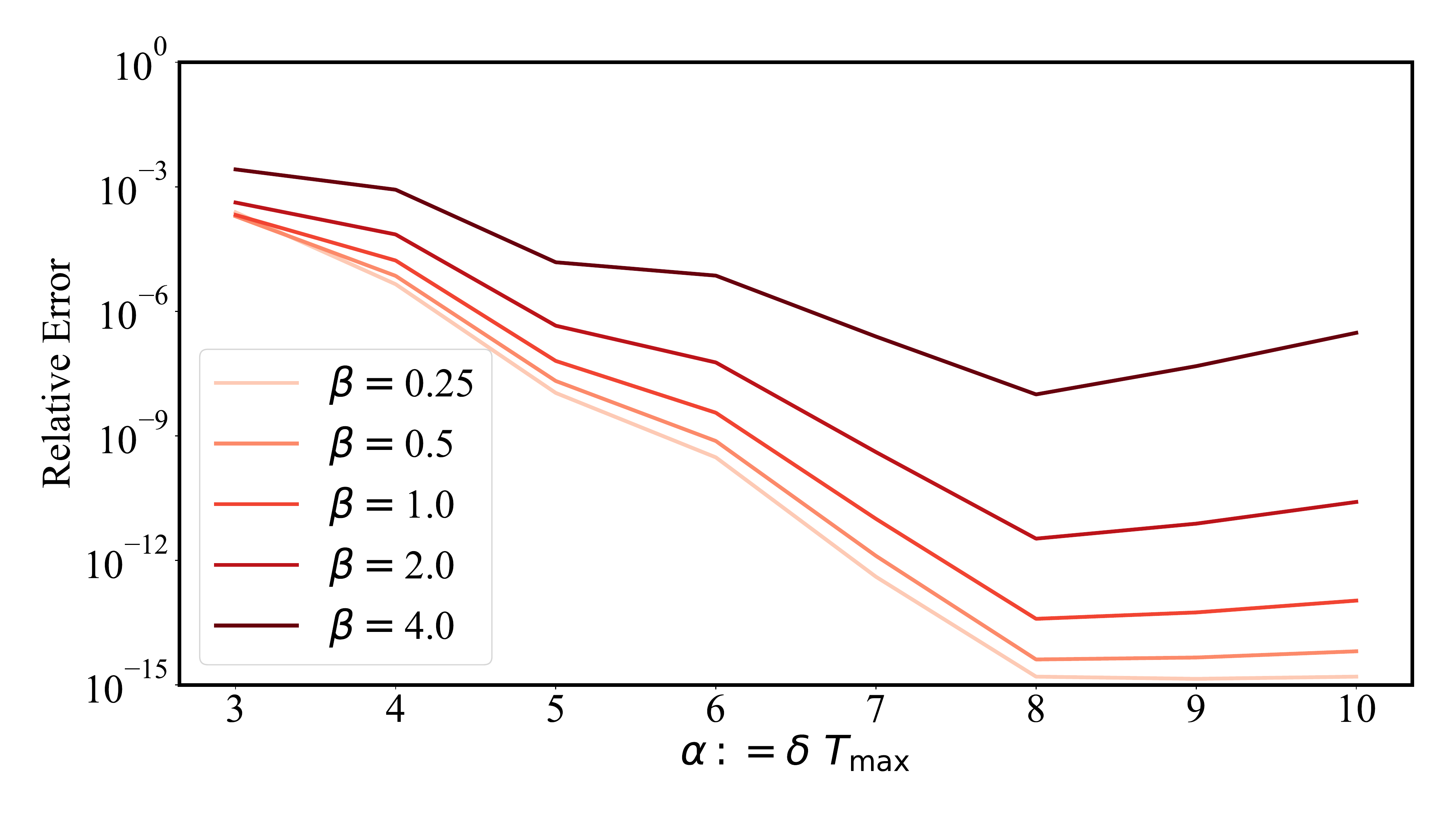}
		\label{fig:alpha_dependece_dp}
		
	}\\
	\subfloat[Single Precision]
	{
		\includegraphics[width=\columnwidth]{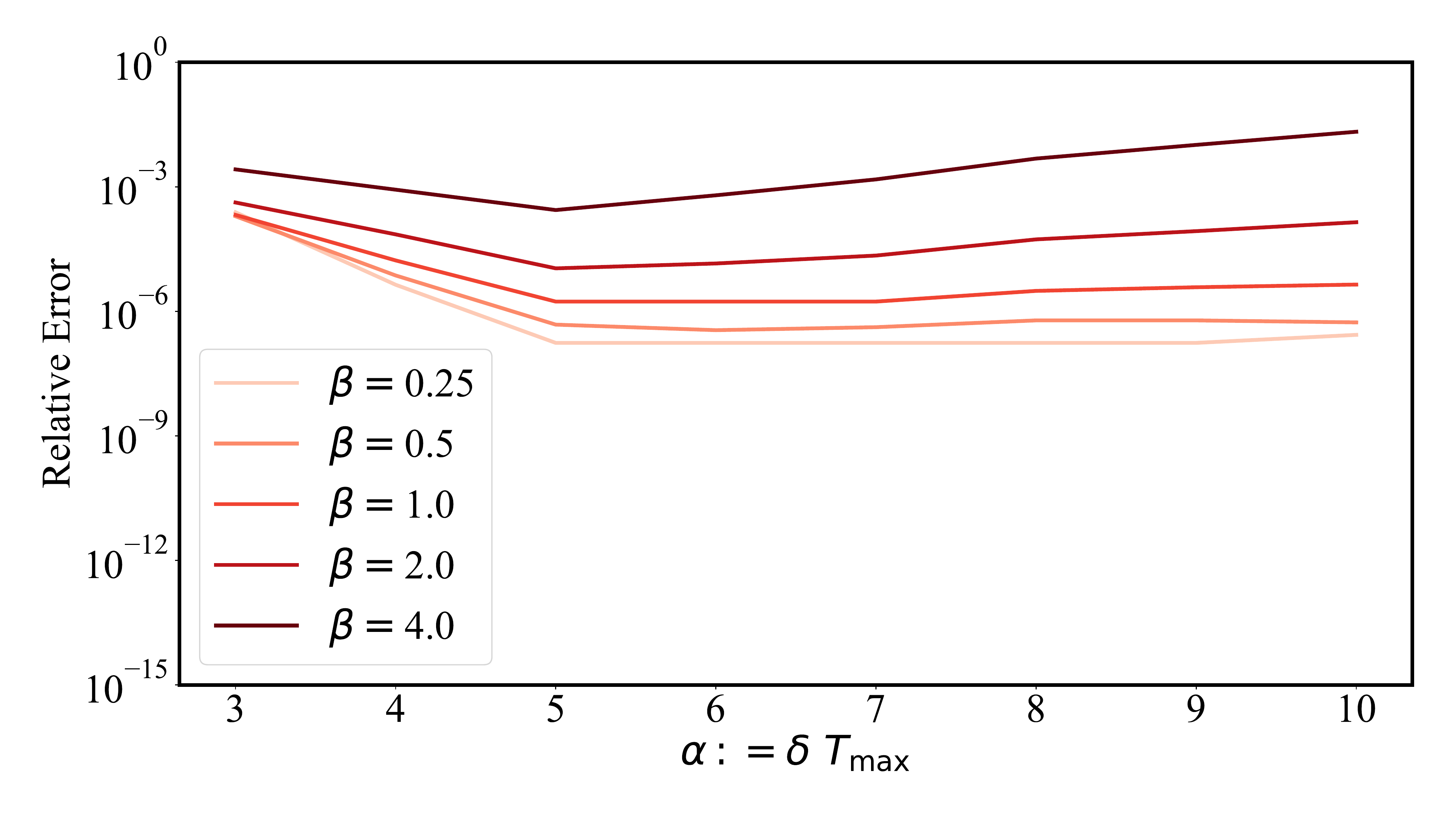}
		\label{fig:alpha_dependece_sp}
	}
	\caption{
		Dependence of the relative error on the Boltzmann weights of the Gaussian filter method on the product $\delta T_\text{max}$ for the time series of Fig.~\ref{fig:spectra}.
		The time series was sampled at a rate $1/\Delta t = 64/\pi$ up to maximum time $T_\text{max} = 4\pi$.
		Calculations were performed using double-precision (top panel) and single-precision (bottom panel) floating-point representations.
	}
\end{figure}

The second limitation is the precision of floating-point arithmetic. 
For small enough $1/\delta$, the tail of the Gaussian filter becomes numerically zero, such that the filtered time series $G_{\psi, \delta}$ is effectively truncated even further instead of being reweighted.
Therefore, a limit to $\alpha \coloneqq \delta T_\text{max}$ is determined by the floating-point precision.
To show this effect,  we plot in Fig.~\ref{fig:alpha_dependece_dp} the relative error in estimating Boltzmann weights as a function of $\delta$ for the time series of Fig.~\ref{fig:spectra}.
The error decays exponentially with increasing $\delta$ up to a point $\alpha = 8$, after which the error plateaus or increases (depending on temperature).
In Fig.~\ref{fig:alpha_dependece_sp}, we show the exact same calculation with single-precision floating numbers.
In this case, the optimal value of $\alpha$ reduces to $\alpha = 5$.
The increase in the error after that point is attributed to the aforementioned reduction in the effective maximum time $T_\text{max}$.
%Therefore, assuming that  the sampling frequency $1/\Delta t$ is large enough, we recommend fixing  $\alpha$ to one of the above values depending on the used floating-point precision.

\begin{figure}[t]
	\centering
	\subfloat[Inverse of Maximum Time]
	{
		\includegraphics[width=\columnwidth]{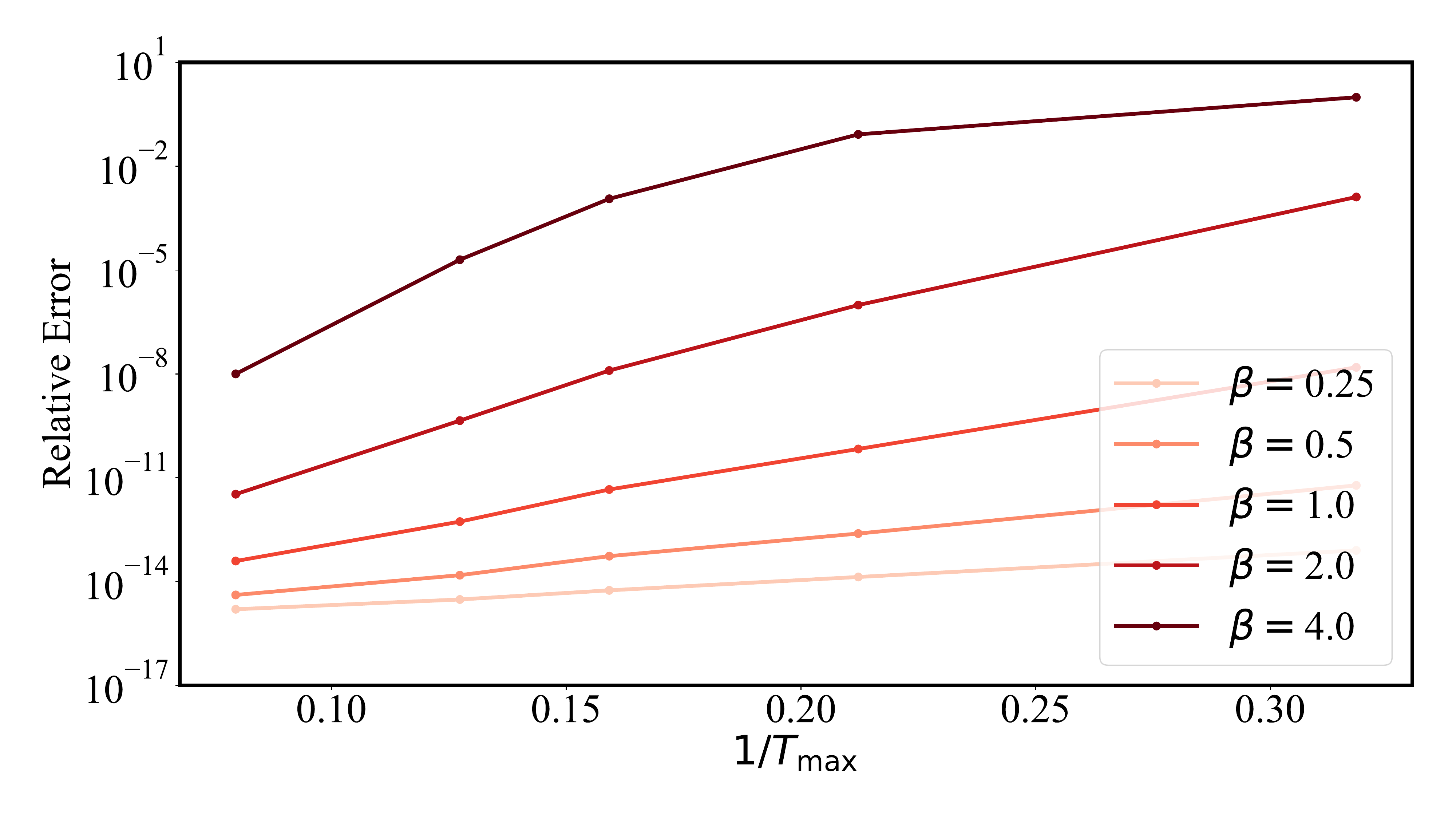}
		\label{fig:T_dependence}
		}\\
	\subfloat[Inverse Temperature]
	{
		\includegraphics[width=\columnwidth]{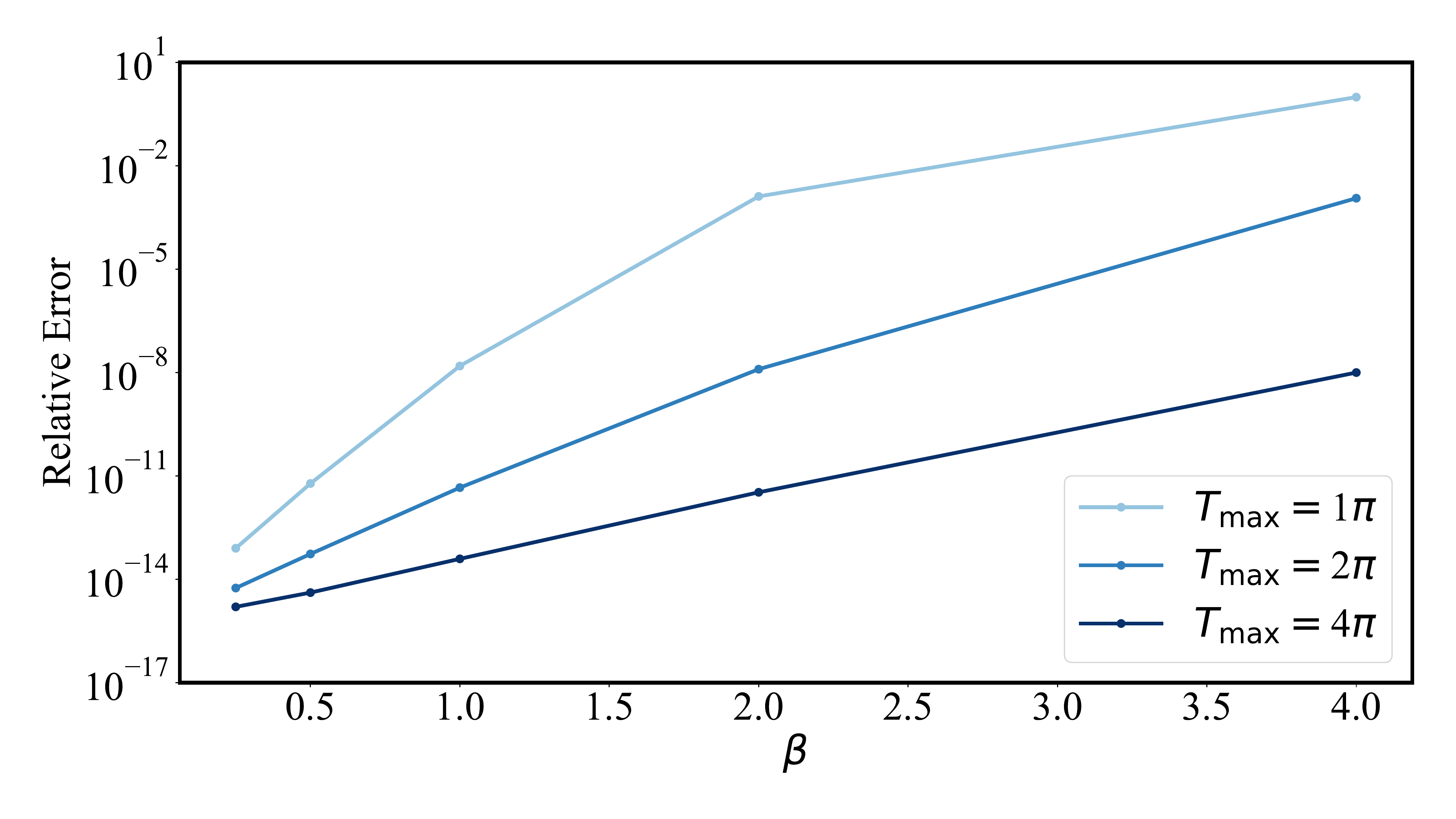}
		\label{fig:beta_dependence}
	}
	\caption{
		Dependence of the relative error of the Gaussian filter method for estimating the Boltzmann weights on the maximum time $T_\text{max}$ (top panel) and inverse temperature $\beta$ (bottom panel) for the time series of Fig.~\ref{fig:spectra}.
		The calculations were performed using double precision.
		The time series were sampled at a rate $1/\Delta t = 64/\pi$, and the filter parameter $\delta$ was chosen such that $\alpha = \delta T_\text{max}= 8$.
		%For performing the integrals, we used a frequency grid with $8192$ points in the interval $[-{\pi}/{\Delta t}, +{\pi}/{\Delta t}[$.
	}
	\label{fig:T_beta_dependence}
\end{figure}

The dependence of the relative error on $T_\text{max}$ while fixing $\alpha$ is shown in Fig.~\ref{fig:T_dependence}.
From this plot, we see that the error decreases exponentially with increasing $1/T_\text{max}$ until it plateaus when hitting numerical accuracy.
Finally, in Fig.~\ref{fig:beta_dependence}, we show the error as a function of inverse temperature $\beta$, which exhibits a similar exponential increase with $\beta$.
The dependence of the error on the different parameters can be summarized in the limit of large $\alpha$  as follows:
\begin{equation}\label{eq:error_scaling}
	\text{Rel. Err.} \sim  \mathcal{O}\left(
 \frac{\exp \left(-\alpha^2\left[ C- \beta/T_\text{max}\right]^2/2 \right)}{\sqrt{2 \pi}\ \alpha  \left[C - \beta/T_\text{max}  \right]} \right)
\end{equation}
where $C$ is some positive constant.
This expression, derived from some simplified assumptions in Appendix~\ref{app:error_scaling}, is only valid when $\beta /T_\text{max}<C$ and matches the observed behavior in the previous plots.
An interesting observation is the duality between temperature and maximum time, where lowering the temperature requires increasing the maximum time proportionally to maintain the same level of accuracy.
Another important observation is that using a fixed $\alpha = \delta T$, the error does not vanish even in the limits $T_\text{max} \to \infty$ or $\beta \to 0$.
However, by choosing the appropriate value of $\alpha$, values close to the numerical accuracy can be reached, as evident from Figs.~\ref{fig:T_dependence} and \ref{fig:beta_dependence}.

\begin{figure}[t]
	\centering
	\subfloat[Inverse of Maximum Time]
	{
		\includegraphics[width=\columnwidth]{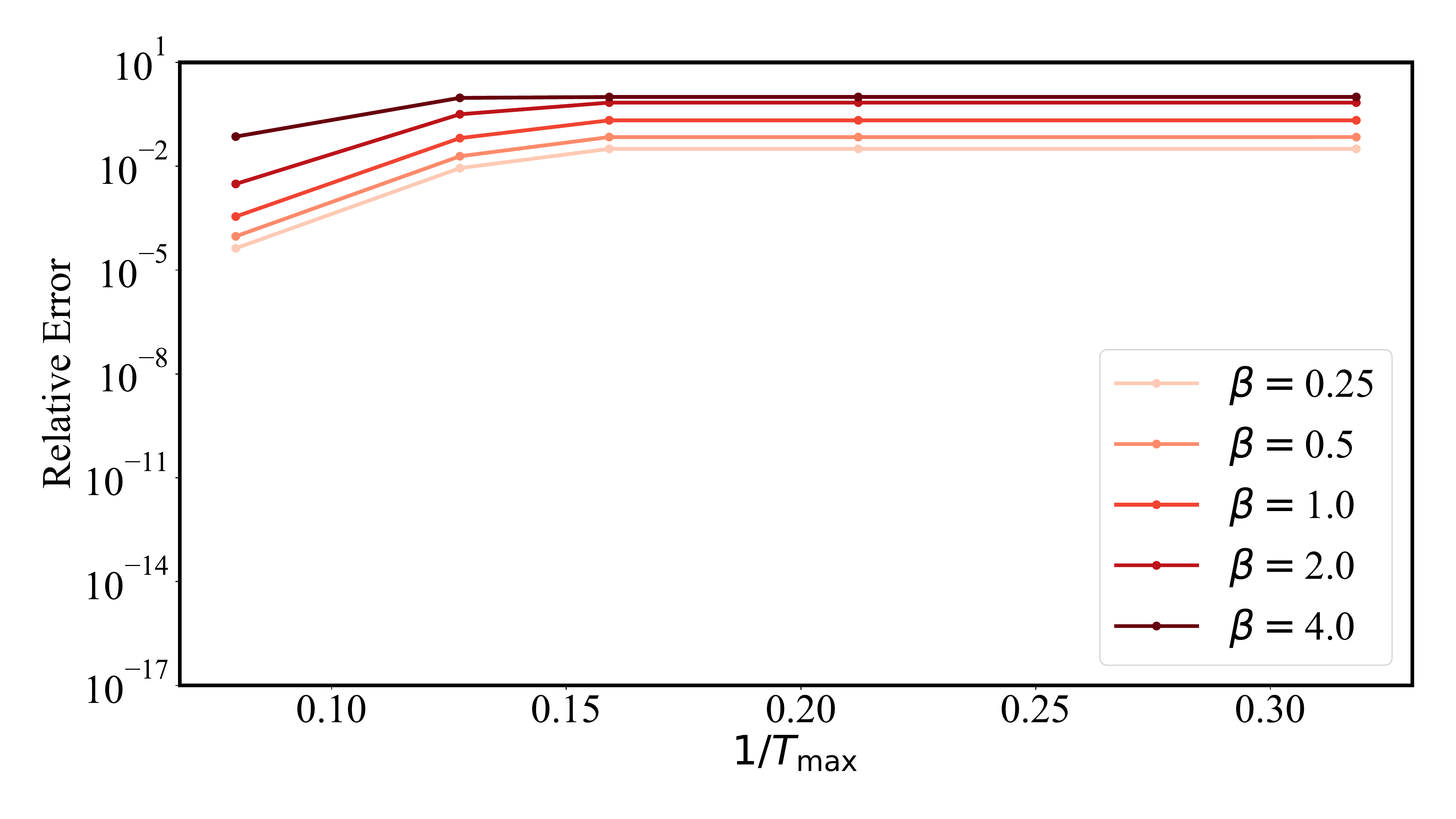}
		\label{fig:T_dependence_low}
	}\\
	\subfloat[Inverse Temperature]
	{
		\includegraphics[width=\columnwidth]{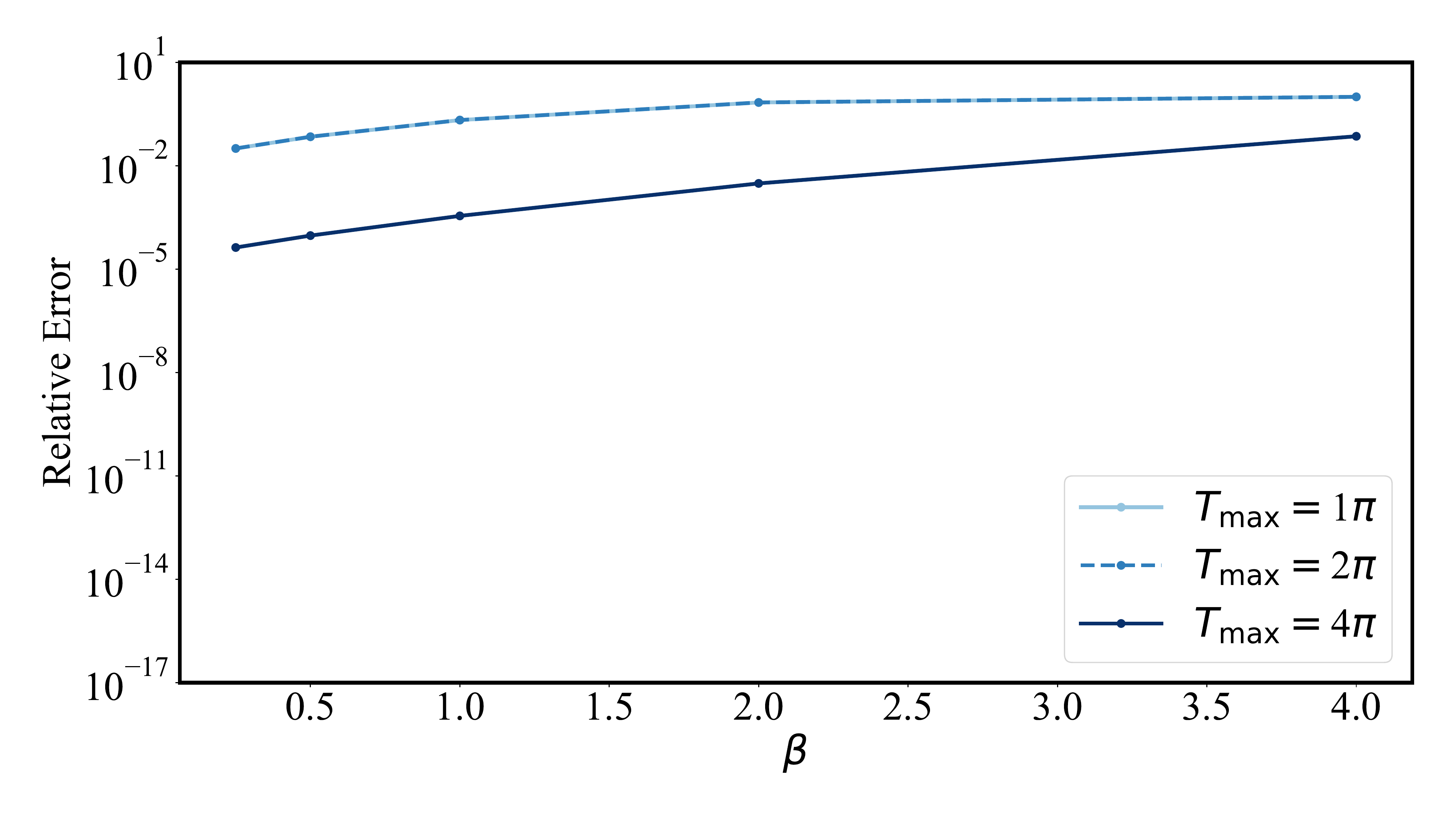}
		\label{fig:beta_dependence_low}
	}
	\caption{
		Same calculations as Fig.~\ref{fig:T_beta_dependence} but using time series sampled at the lower rate $1/\Delta t = 16/\pi$. To limit the effect of this low sampling rate on the broadened density (cf. Fig.~\ref{fig:interference}), a cap is put on the filter parameter: $\delta_\text{max} = (\pi/\Delta t-E_0)/2$, where $E_0=-13.477(8)$ is the ground state energy.
        Note that curves of $T_\text{max} = \pi$ and $T_\text{max} = 2 \pi$ are on top of each other.
	}
	\label{fig:T_beta_dependence_low}
\end{figure}

It should be emphasized that the results discussed so far used a sampling frequency much higher than the Nyquist frequency in order to minimize the aforementioned interference effects on the broadened density.
When the sampling frequency is relatively low, we cannot set $\delta$ to its optimal value $T_\text{max}/\alpha$.
Instead, it should be chosen such that $\delta \ll \pi/\Delta t - \text{max} \left\{ |E_i| \right\}$, which increases the error significantly as illustrated in Fig.~\ref{fig:T_beta_dependence_low}.
In practice, the need for high sampling frequency puts yet an additional burden on the quantum resources required for an accurate reconstruction of Boltzmann weights.

\subsection{Non-Negative Least Squares}
In the previous section, we have shown how to systematically reduce the error of the Gaussian filter method by increasing both the maximum time and the sampling frequency.
When either is low, however, it remains problematic to extract Boltzmann weights, especially at relatively low temperatures.

In the following, we introduce the main contribution of our work:  a method for obtaining much better results using the same limited data set. The critical idea is using the fact that physical densities of states must be non-negative.
So instead of taking the time series and directly computing its Fourier transform, we look among the non-negative densities for the one whose inverse Fourier transform best fits the time series.
This can be found using the non-negative least squares (NNLS) method, written formally as:
\begin{equation}
	D_\psi^\text{NNLS}(\omega) = \underset{D_\psi(\omega)\geq 0 }{\argmin}\ \chi^2[D_\psi(\omega)].
\end{equation}
where $\chi^2$ measures how well the density fits the time series:
\begin{equation}\label{eq:chi}
	\chi^2[D_\psi(\omega)] \coloneqq  \sum_{i=1}^{n_t} \left|G(t_i) - \int d\omega\ e^{-i \omega t_i} D_\psi(\omega) \right|^2,
\end{equation}
where $n_t$ is the number of time points.
In practice, we construct a frequency grid in the interval $[-\pi /\Delta t, +\pi/\Delta t[$ with enough points to resolve the details of the density, and we use it to build the inverse Fourier transform matrix $M_{i, j} = e^{-  i\omega_j t_i}$.
Then we solve the linear system $\vec{G} = \vec{M} \vec{D}$ under the constraint $\vec{D}\geq 0$, where $\vec{G}$ and $\vec{D}$ are vector representations of the time series and the density, respectively.
The solution is obtained numerically using the active set algorithm by Lawson and Hanson~\cite{Lawson95}.

\begin{figure}[t]
	\includegraphics[width=\columnwidth]{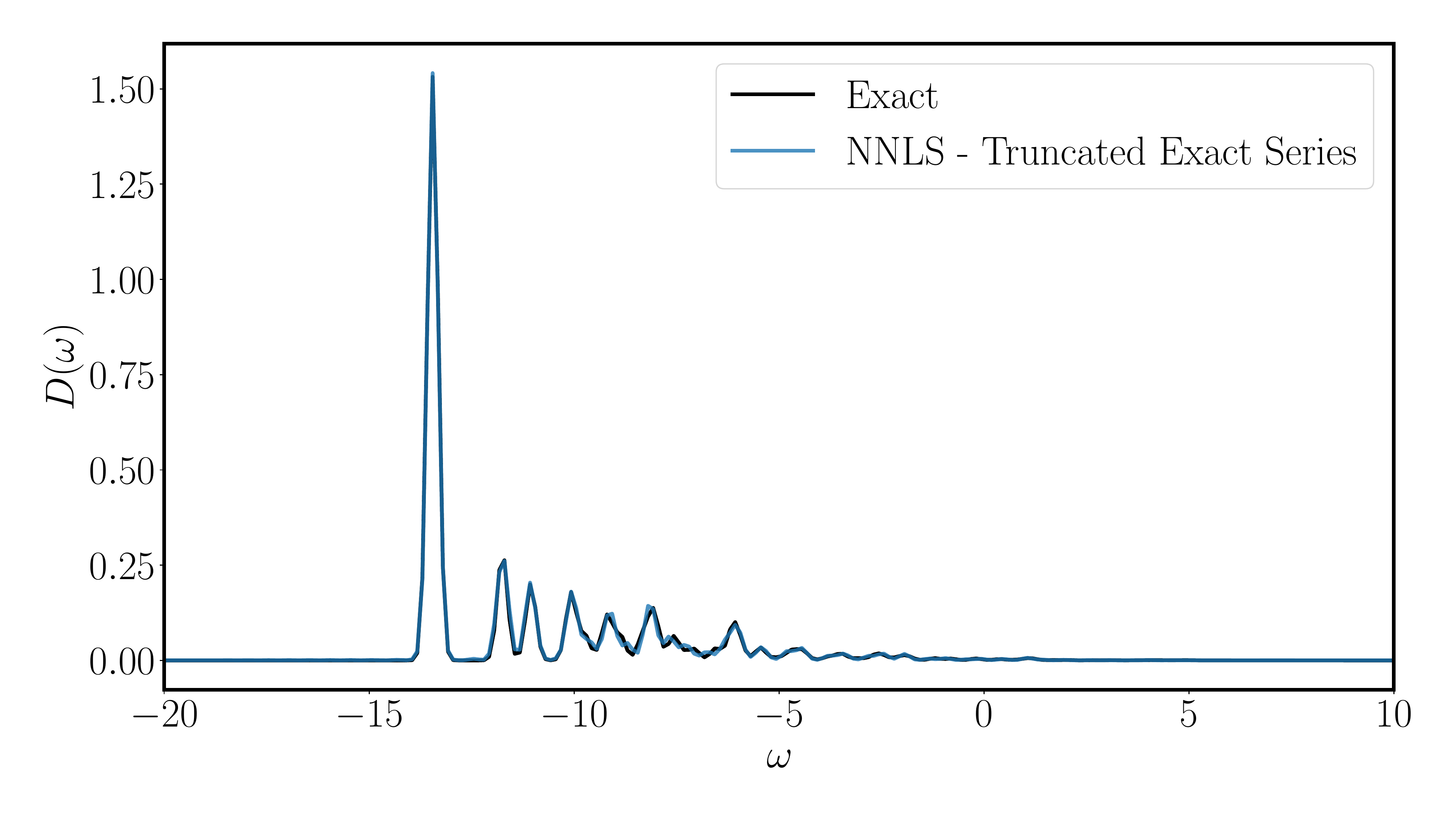}
	\caption{\label{fig:spectra_nnls} 
		Fourier transform of the time series of Fig.~\ref{fig:spectra} using non-negative least squares method.
		For ease of comparison, both the exact density and NNLS are convoluted with a Gaussian whose width equals the grid spacing $\Delta \omega = 1/8$.		
	}
\end{figure}

\begin{figure}[t]
	\centering
	\subfloat[Inverse of Maximum Time]
	{
		\includegraphics[width=\columnwidth]{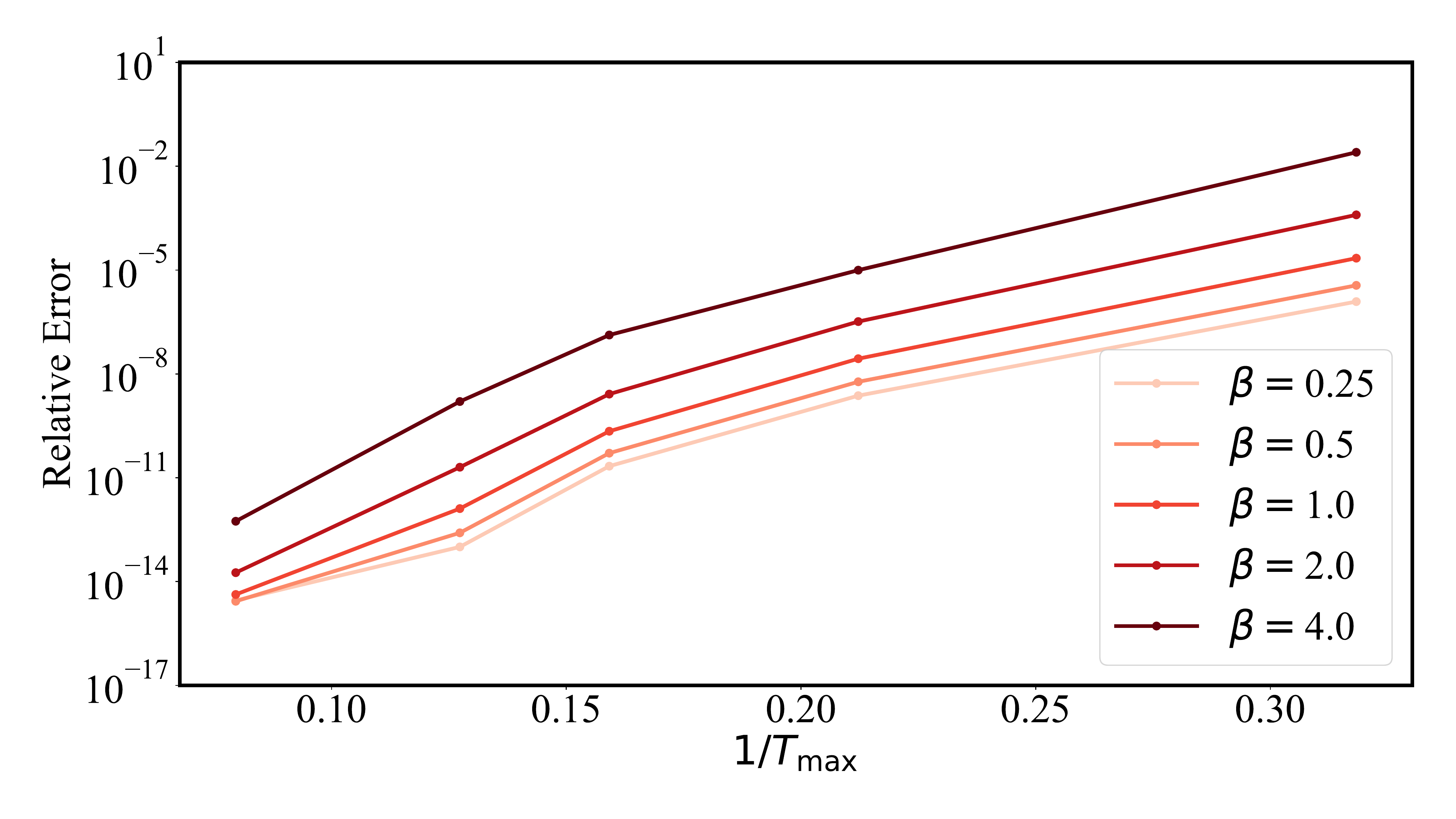}
		\label{fig:T_dependence_nnls}
	}\\
	\subfloat[Inverse Temperature]
	{
		\includegraphics[width=\columnwidth]{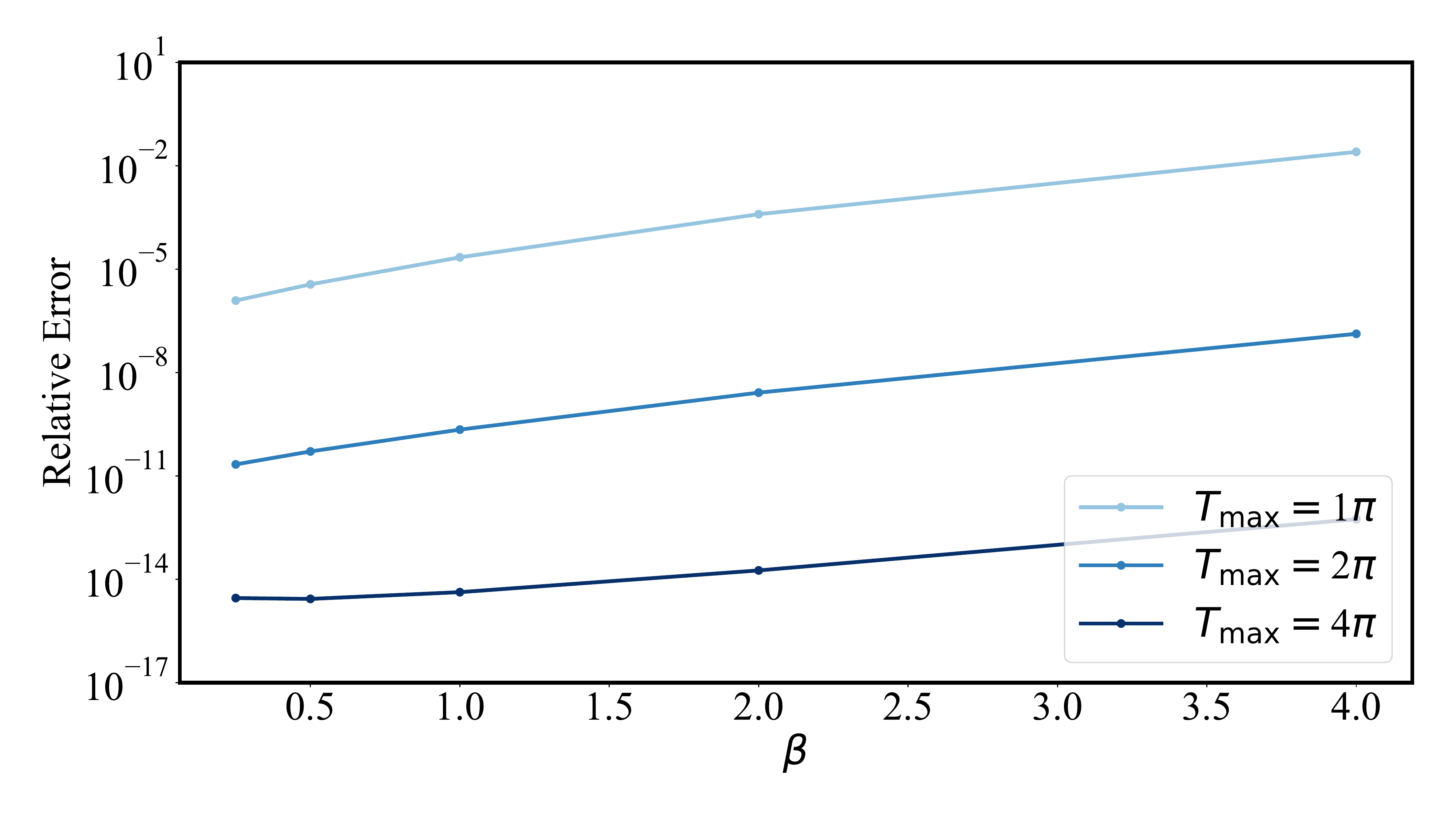}
		\label{fig:beta_dependence_nnls}
	}
	\caption{
		Dependence of the relative error of the non-negative least squares method on the maximum time $T_\text{max}$ (top panel) and inverse temperature $\beta$ (bottom panel) for the time series of Fig.~\ref{fig:spectra}.
		The time series were sampled at a rate $1/\Delta t = 16/\pi$ as in Fig.~\ref{fig:T_beta_dependence_low}.
	}
	\label{fig:T_beta_dependence_nnls}
\end{figure}

In Fig.~\ref{fig:spectra_nnls}, we show the NNLS solution of the time series in Fig.~\ref{fig:spectra}, and the result matches the exact density on the specified $\omega$ grid.
This is a significant improvement over the direct Fourier transform and translates into several orders of magnitude reduction in the error when calculating Boltzmann weights.
In Fig.~\ref{fig:T_beta_dependence_nnls}, we show the relative error of NNLS as a function of maximum time and temperature.
These results were obtained using the lower sampling rate $1/\Delta t = 16/\pi$ and are superior to the ones obtained from the Gaussian filter method using the same data (cf. Fig.~\ref{fig:T_beta_dependence_low}).

When obtaining Boltzmann weights from NNLS densities, we recommend to additionally use a very mild Gaussian filter of width proportional to the spacing of the frequency grid.
Without it, NNLS has to choose between two neighboring frequencies when placing a delta function (or a very narrow peak).
This makes low-temperature results sensitive to how close one of the grid points is to the exact location of the delta function.
Using the suggested filter, the broadened density is easier to represent on the discretized grid.
It should be emphasized that, unlike in the original Gaussian filter method, the role of the broadening is only to regularize the effect of frequency discretization and can be made arbitrarily small using a finer $\omega$ grid.

\subsubsection{Influence of shot noise on NNLS: Quantile Filter and Discrepancy Principle}
\begin{figure}[t]
	\includegraphics[width=\columnwidth]{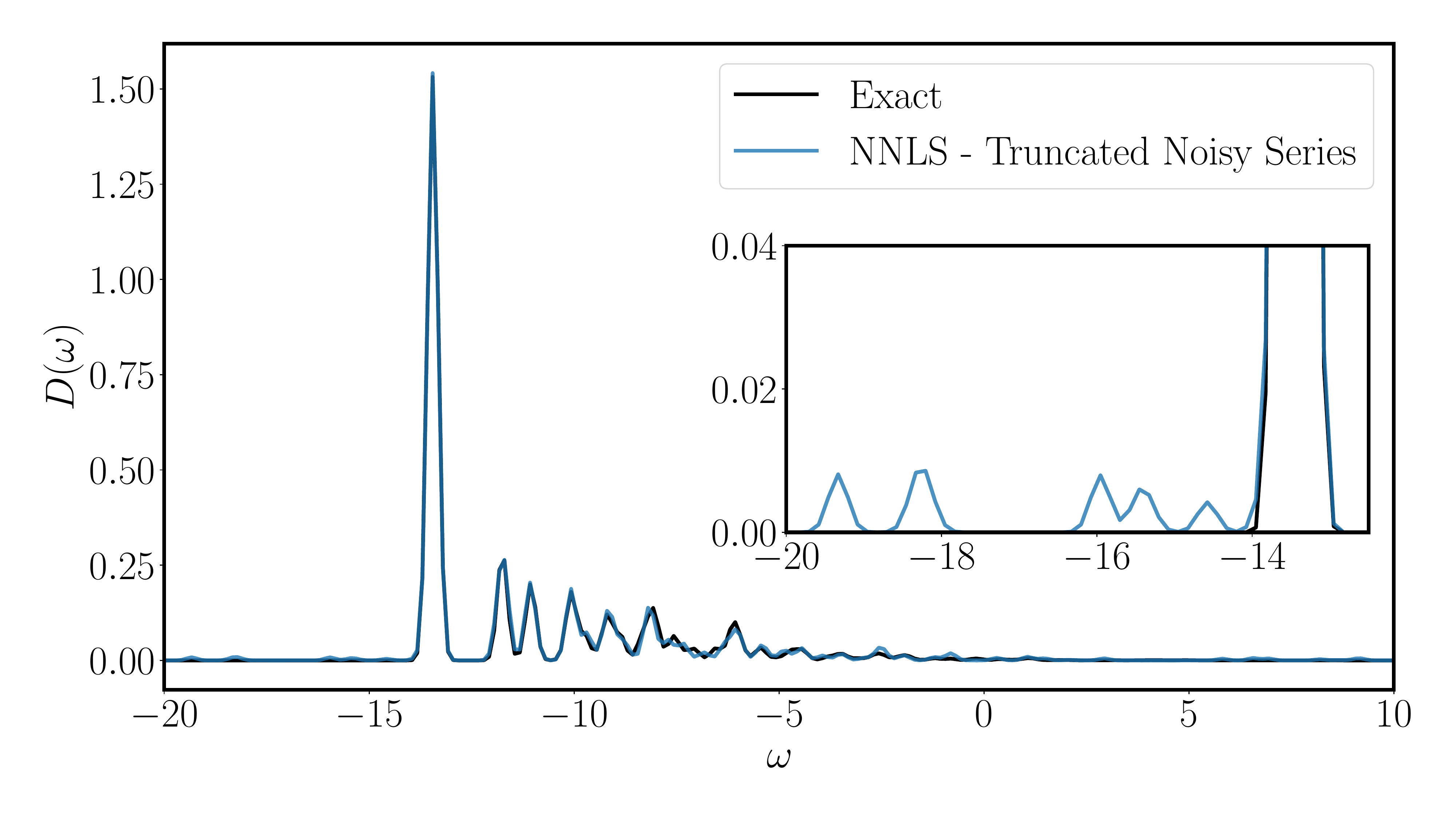}
	\caption{\label{fig:spectra_nnls_noisy} 
		Fourier transform of a \emph{noisy} version of the time series of Fig.~\ref{fig:spectra} using non-negative least squares method.
    The shot noise is approximated as Gaussian noise of zero mean and variances given by Eq.~\eqref{eq:variance}.
		Here $N_\text{shots} = 1000$ is used.
	}
\end{figure}

The discussion so far has been about limited but numerically exact data.
Using quantum computers, however, the time series are estimated by averaging over a finite number of shots from the quantum circuits.
As a result, the time series contains statistical noise that scales as $1/\sqrt{N_\text{shots}}$.
In particular, the shot noise on the real and imaginary parts of Loschmidt echo $G$ have zero mean and variances 
\begin{align}\label{eq:variance}
\begin{split}
    \operatorname{Var}[\Re{(G)}] &= \frac{[1-\Re{(G)}^2]}{N_\text{shots}}\;, \\
    \operatorname{Var}[\Im{(G)}] &= \frac{[1-\Im{(G)}^2]}{N_\text{shots}}\;.
    \end{split}
\end{align}
The effect of this noise on the NNLS density is introducing small, randomly placed peaks with a total weight proportional to the noise level as illustrated in Fig~\ref{fig:spectra_nnls_noisy}.
When these spurious peaks occur at a frequency lower than the first true peak, they introduce a disproportionally larger error in the Boltzmann weights, especially at low temperatures.

The first step in amending the effect of noise is redefining the fit function $\chi^2$ of NNLS in Eq.~\eqref{eq:chi} such that each term is weighted by the inverse of its variance:
\begin{equation}\label{eq:chi_weighted}
\begin{split}
	\chi^2[D_\psi(\omega)] \coloneqq  &\sum_{i=1}^{n_t} \frac{\Re\left[G(t_i) - \int d\omega\ e^{-i \omega t_i} D_\psi(\omega)\right]^2}{\operatorname{Var}[\Re(G(t_i)]} \\
 +& \sum_{i=1}^{n_t} \frac{\Im\left[G(t_i) - \int d\omega\ e^{-i \omega t_i} D_\psi(\omega)\right]^2}{\operatorname{Var}[\Im(G(t_i)]}.
 \end{split}
\end{equation}
This drives NNLS to fit accurate data points better than noisier ones.
For example, Loschmidt echo at $t=0$ is known exactly and corresponds to the normalization constraint $G(0) = \int d\omega D_\psi(\omega) = 1$. This constraint can then be approximately enforced during NNLS by using a relatively large weight for $G(0)$ terms in the fit function.

To limit the effect of the noise on the Boltzmann weights, we introduce, as a post-processing step, a quantile filter where the density outside a specific quantile range $[q, 1-q]$ is truncated.
The quantile of a normalized density function is defined as $Q_\psi(q) = \operatorname{CFD}_\psi^{-1}(q)$,
where $\operatorname{CDF}_\psi(\omega) = \int_{-\infty}^{\omega} d\omega^\prime\ D_\psi(\omega^\prime)$ is its cumulative distribution function (CDF).
If value of $q$ is chosen appropriately, leading and trailing spurious peaks will be eliminated. 
The main advantage of the quantile filter, as opposed to the simple cutting procedure used by the Gaussian filter, is that the proper value of $q$ is relatively stable regarding the broadening of the density.

The optimal truncation value $q$ can be determined automatically using the discrepancy principle~\cite{Groetsch84, Morozov84}.
According to this principle, the truncation level is chosen such that the truncated density $\tilde{D}^q_\psi(\omega)$ fits the time series up to the expected amount of noise on that series.
When $N_\text{shots}$ is large, the shots noise is approximately Gaussian, and thus the reweighted fit function of Eq.~\eqref{eq:chi_weighted} follows the chi-squared distribution with $2 n_t$ degrees of freedom, where $n_t$ is the number of time points.
The mean value of this distribution equals its degrees of freedom, so the discrepancy principle suggests choosing the quantile value $q$ such that $\chi^2[\tilde{D}^q_\psi(\omega)] = 2 n_t$.
A larger value can be used to avoid accidental over-fitting of the noise or if other sources of systematic error are expected (e.g., Trotter error).

\begin{figure}[t]
	\centering
	\includegraphics[width=\columnwidth]{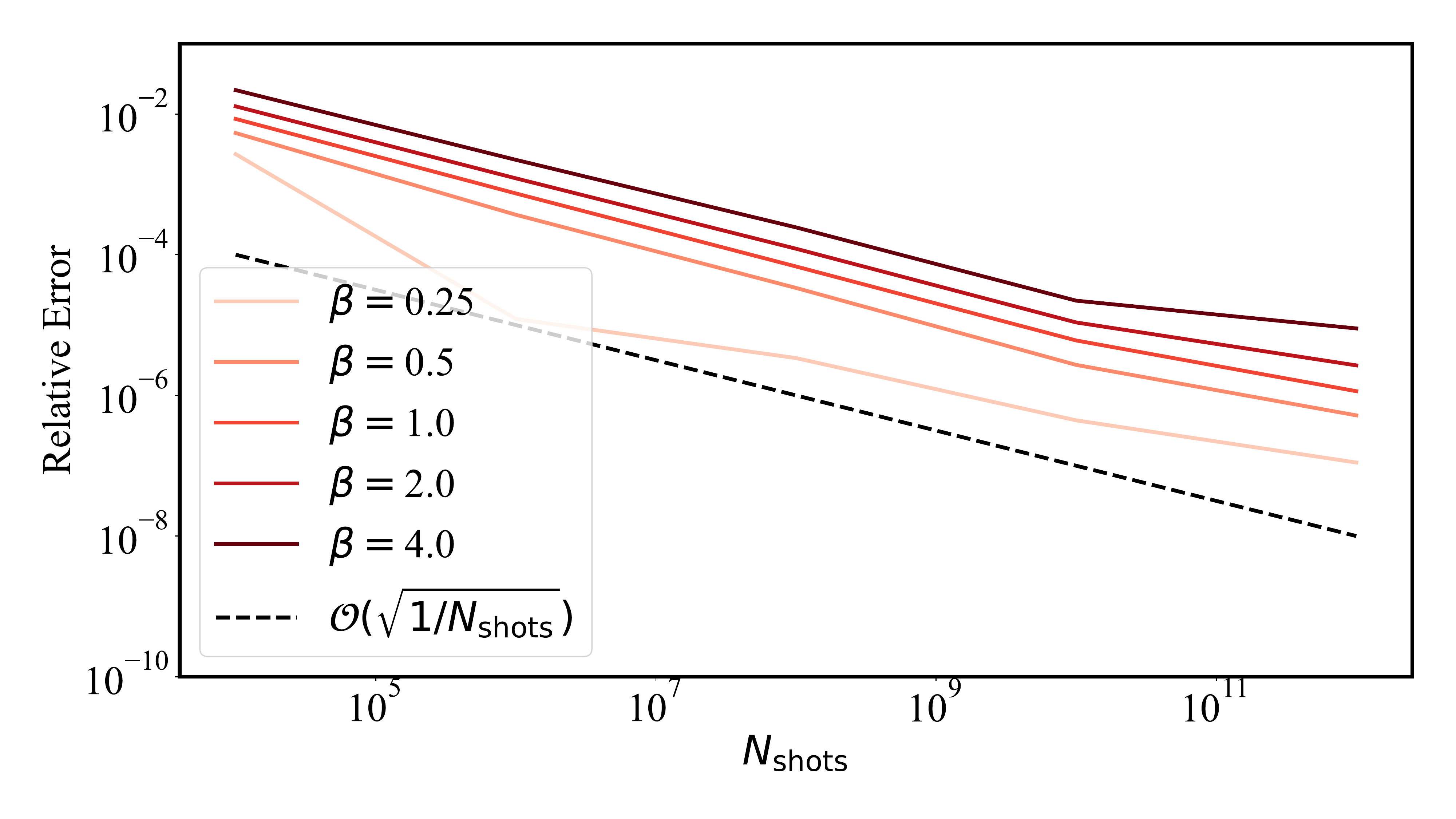}
	\caption{
		Dependence of the relative error of the  non-negative least squares method on the number of shots for the time series of Fig.~\ref{fig:spectra}.
		The time series were sampled at a rate $1/\Delta t = 16/\pi$ with maximum time $T_\text{max}=4\pi$.
	}
	\label{fig:noise_dependence}
\end{figure}

In Fig.~\ref{fig:noise_dependence}, we plot the relative error in Boltzmann weights as a function of the number of shots for the time series of Fig.~\ref{fig:spectra}.
We simulated the effect of shot noise using Gaussian random variables of appropriate variances given by Eq.~\eqref{eq:variance}.
The density was then reconstructed using NNLS and truncated with a quantile filter, where the truncation level was determined automatically using the discrepancy principle.
The plot shows that using the quantile filter in combination with the discrepancy principle effectively regularizes the impact of shot noise and allows systematically improvable results.
The error on the Boltzmann weights scales linearly with the error on the time series, and thus it scales as $1/\sqrt{N_\text{shots}}$. In particular, no additional hyperparameter is introduced as the algorithm only takes the number of shots as an additional input parameter.

\subsubsection{Benchmarking NNLS}
To test the overall effectiveness of our Wick's rotation strategy (NNLS + quantile filtering + discrepancy principle), we calculate the exact Boltzmann weights $W^\star_\psi$  for all the states of the 10-sites TFIM and compare them with the weights $W_\psi$ estimated via Wick's rotation of the time series of these states.
As a global measure of the error made during Wick's rotation, we need a statistical distance quantifying how different the estimated Boltzmann distribution $p_\psi \coloneqq {W_\psi}/{\sum_{\psi^\prime} W_{\psi^\prime}}$ is from the exact one $p_\psi^\star \coloneqq {W^\star_\psi}/{\sum_{\psi^\prime} W^\star_{\psi^\prime}}$.
One option, which we employ here, is the relative entropy (also known as Kullback–Leibler divergence) of the estimated Boltzmann distribution with respect to the exact one
\begin{equation}
	\operatorname{KL}(p||p^\star) = \sum_\psi p_\psi \log\left(\frac{p_\psi}{p^\star_\psi}\right)\;.
\end{equation}
It is worth noting that we also checked the $L_1$-distance (also known as Kolmogorov distance) and it shows similar trends to the entropy, so we restrict the discussion below to the later.

\begin{figure}[t]
	\centering
            \includegraphics[width=\columnwidth]{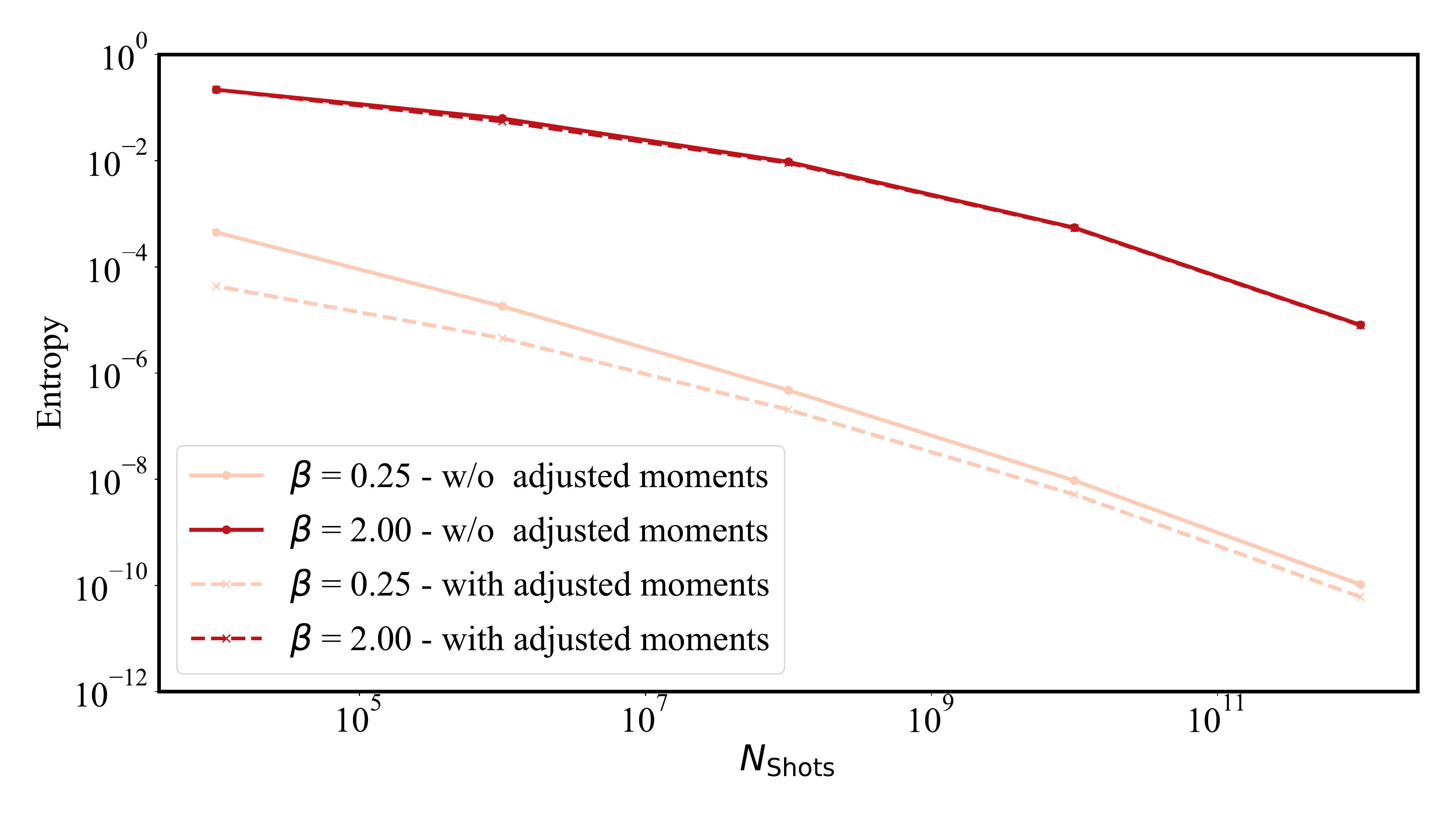}
	\caption{
		Dependence of the relative entropy error on the number of shots for 10-sites TFIM with $h_x=1$.
		The time series were sampled at a rate $1/\Delta t = 16/\pi$ with maximum time $T_\text{max}=4\pi$.
	}
	\label{fig:full_noise_dependence}
\end{figure}

In Fig.~\ref{fig:full_noise_dependence}, we plot the error for different values of $N_\text{shots}$ at $h_x=1$ and different $\beta$. We find a slight improvement of these results when enforcing constraints on the first few moments of the density, in this case, the mean energy $E_\psi \coloneqq  \braket{\psi| H | \psi} = \int d\omega D_\psi(\omega) \omega$, and the energy variance  $\braket{\psi| (H-E_\psi)^2| \psi} = \int d\omega D_\psi(\omega) (\omega-E_\psi)^2$.
These moments can be calculated efficiently and accurately on the classical computer for product states.
The truncation performed by the quantile filter will generally violate these constraints. However, they can be reintroduced by shifting and rescaling the truncated density to fix its moments to their exact values.
The plot shows the relative error in Boltzmann weights using adjusted moments which makes a slight but noticeable improvement for low numbers of shots and high temperatures.

\begin{figure}[t]
	\centering
	\includegraphics[width=\columnwidth]{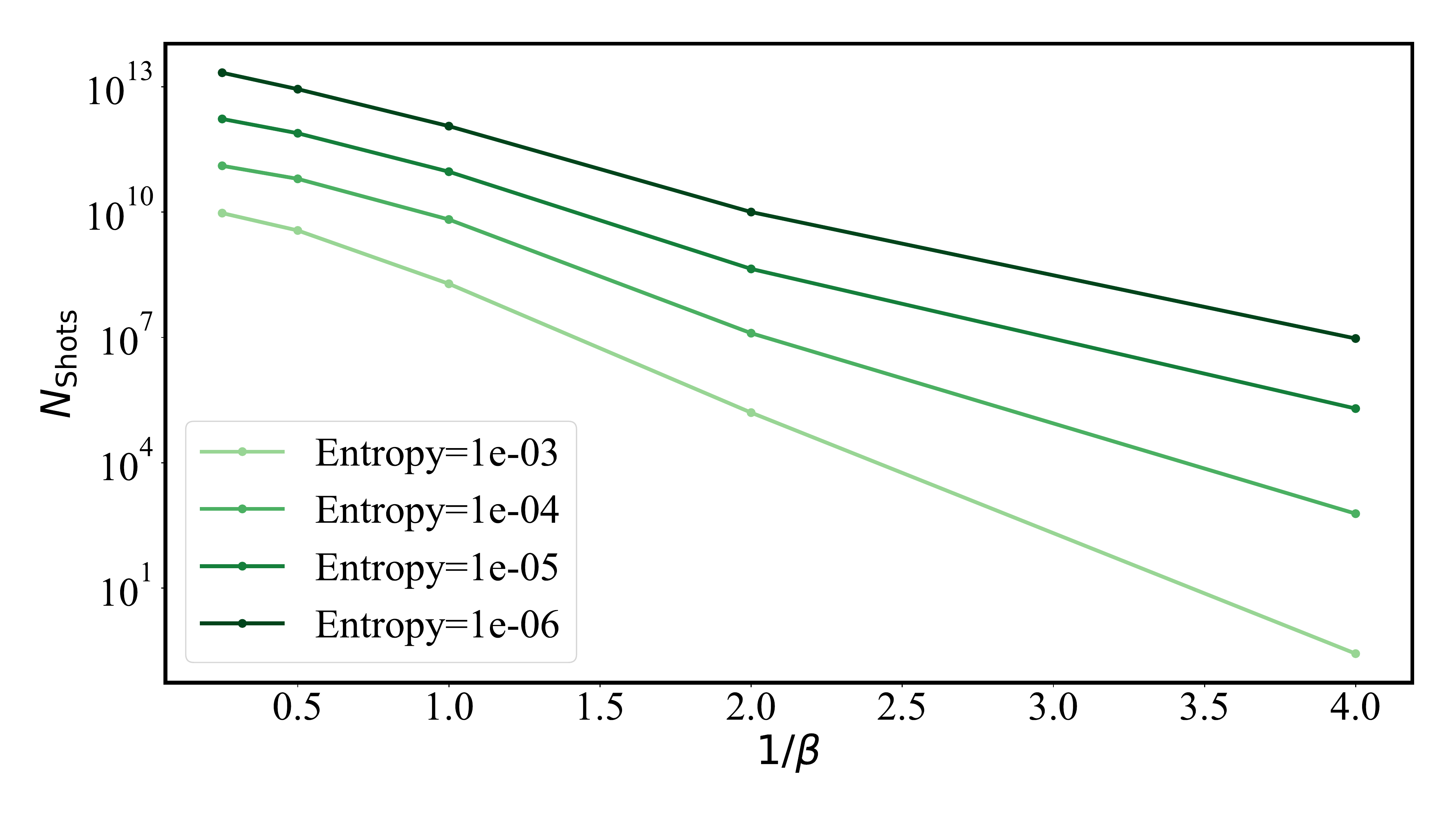}
	\caption{
		Temperature dependence of the number of shots needed for achieving the labeled relative entropy for 10-sites TFIM with $h_x=1$.
		The time series were sampled at a rate $1/\Delta t = 16/\pi$ with maximum time $T_\text{max}=4\pi$.
	}
	\label{fig:shots_beta_dependece}
\end{figure}

Fig.~\ref{fig:shots_beta_dependece} shows the number of shots required to reach a specific level of accuracy in the Boltzmann distribution for different temperatures.
The number of shots scales exponentially with the inverse temperature.
This exponential scaling comes from the fact that errors in the estimated density gets amplified by the Boltzmann factor $e^{-\beta \omega}$ which itself scales exponentially with the inverse temperature.

\begin{figure}[t]
	\centering
	\includegraphics[width=\columnwidth]{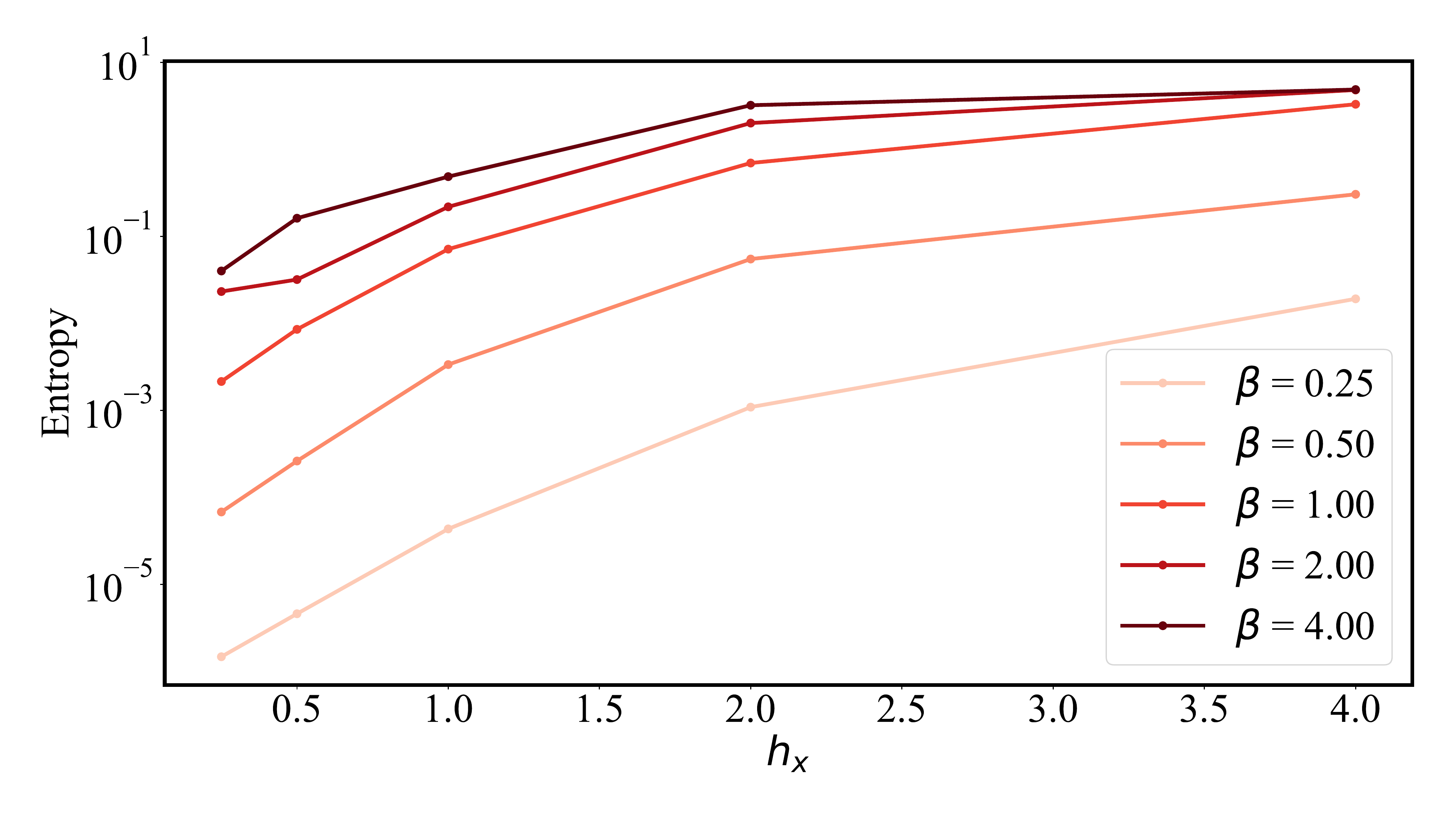}
	\caption{
		Dependence of the relative entropy error on the strength of magnetic field for 10-sites TFIM with $N_\text{shots}=10000$.
		The time series of $h_x=0.25, 0.5, 1, 2, 4$ were sampled at rates $1/\Delta t = 12\pi, 12\pi, 16/\pi, 24/\pi, 48/\pi$, respectively, with maximum time $T_\text{max}=4\pi$.
		The sampling rates were chosen such that the Nyquist condition of Eq.~\eqref{eq:nyquist} is satisfied.
	}
	\label{fig:hx_dependece}
\end{figure}

In Fig.~\ref{fig:hx_dependece}, we also show the dependence of the relative entropy error on the strength of magnetic field parameter $h_x$.
As $h_x$ increases, Wick's rotation becomes more difficult.
This can be explained by the fact that the density of states gets broader with increasing $h_x$ (the energy variance equals $h_x^2$ times the number of sites).
With wider support, more of the shot noise is located  inside the main body of the density and is not removed by the quantile filter, which cuts only the tails. 
This observation suggests that sampling entangled states (instead of simple product states) with lower energy variances can lower the effect of noise on Boltzmann weights.
Note that in the limit when the sampled states are eigenstates (e.g. $h_x=0$), moments-adjusted densities can become exact even in the presence of shot noise.
This is a result of enforcing the moments of the density by hand and contingent on using exact values of the eigenenergies.
If the moments themselves, however, contain errors, then these errors will be reflected in the Boltzmann weights and scale exponentially in inverse temperature.
For example, let $\psi_1, \psi_2$ be two eigenstates with energies $E_1, E_2$, respectively. 
Then the ratio of their Boltzmann probabilities is $e^{-\beta (E_2 - E_1)}$. If an error $\epsilon$ is made in estimating $E_2-E_1$, then the relative error in the ratio is $|1-e^{-\beta \epsilon}|$, which is exponential in inverse temperature.

\begin{figure}[t]
	\centering
        \includegraphics[width=\columnwidth]{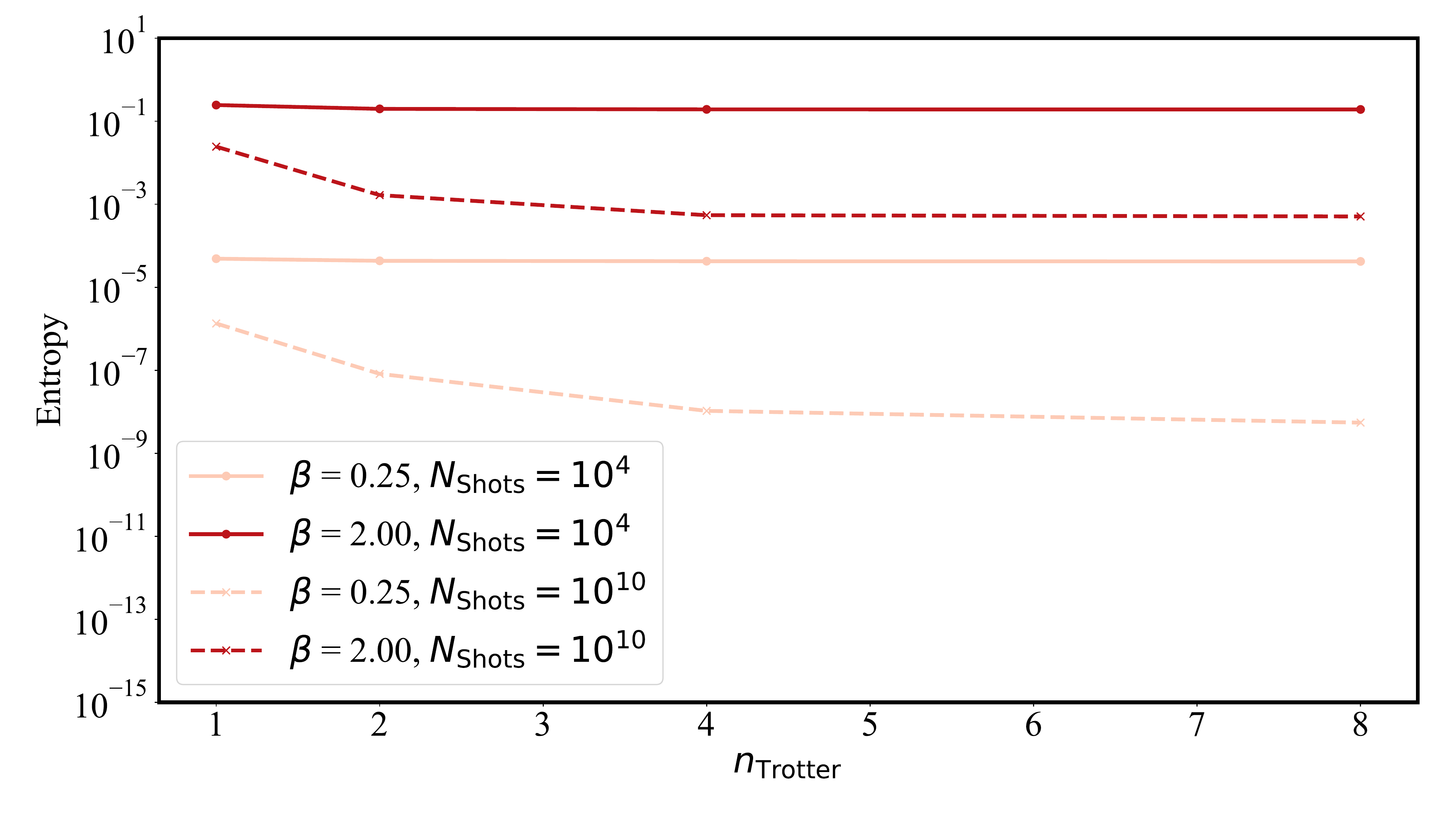}
	\caption{
		Dependence of the relative entropy error on the number of Trotter steps $n_\text{Trotter}$ for 10-sites TFIM with $h_x=1$.
		The time series were sampled at a rate $1/\Delta t = 16/\pi$ with maximum time $T_\text{max}=4\pi$.
		Note that $n_\text{Trotter}$ represents the number of Trotter steps per sampling period $\Delta t$, so that the total number of Trotter steps for time $t$ is $n_\text{Trotter}\times(t/\Delta t)$.
	}
	\label{fig:trotter_dependence}
\end{figure}

Finally, we investigate the effect of Trotterization on Wick's rotation.
We expect that once the Trotter error is below the statistical shot noise, Trotterization has no noticeable effect on Boltzmann weights.
We used a Trotter step $\delta t \coloneqq \Delta t/n_\text{Trotter}$, where $n_\text{Trotter}$ represents the number of Trotter steps per sampling period $\Delta t$.
The results are plotted in Fig.~\ref{fig:trotter_dependence} for different numbers of shots and confirm our expectations.
For a high number of shots, the time series is accurate enough such that increasing the number of Trotter steps systematically decreases the error in Boltzmann weights.
On the other hand, using a relatively low number of shots using one Trotter step per sampling period is already enough.

\section{Simulation Results}
To demonstrate the robustness under the combined effects of Trotter error, shot noise and Monte Carlo sampling errors, we implemented the whole algorithm using quantum circuits simulated via the IBM's AerSimulator with no hardware errors.
We performed calculations for the 16-sites TFIM with $h_x/J=1$ at different temperatures.
Loschmidt echoes were evaluated on $n_t = 8$ times points equally spaced between zero (excluded) and $JT_\text{max}=1$ (included).
We used a variable Trotter step with a maximum value of $0.25$. This implies that the first two time points use one Trotter step, the third and fourth points use two steps and so on. 
We sampled the output of each circuit using $10000$ shots.
With this number of shots, Trotter error is still appreciable compared to the shot noise.
To regulate its effect on the density $D(\omega)$, we applied the discrepancy principle with a higher value $\chi^2[\tilde{D}^q_\psi(\omega)] = 5 n_t$.

We used four Monte Carlo chains for each temperature value, each with 512 samples and a different random seed.
Two chains started from the all-spin-up state, and the other two started from the all-spin-down state.
We discarded the first 32 samples of each chain as a burn-in period. 
To save on the number of executed quantum circuits, we implemented a caching mechanism that allows reusing the shots when the same quantum state is proposed again. 
The savings gained from this trick depends on the length of the Markov chain and the temperature.
At lower temperatures, most samples are low energy states, while at high temperatures, the Boltzmann distribution is more uniform with a lower probability of repeated sampling.
The ratio of unique states for $\beta_c/3$ chains was around 94\%,  for $\beta_c$ was around 63\%, and for $8\beta_c/3$ was around 23\%.

\begin{figure}[t]
	\centering
	\includegraphics[width=\columnwidth]{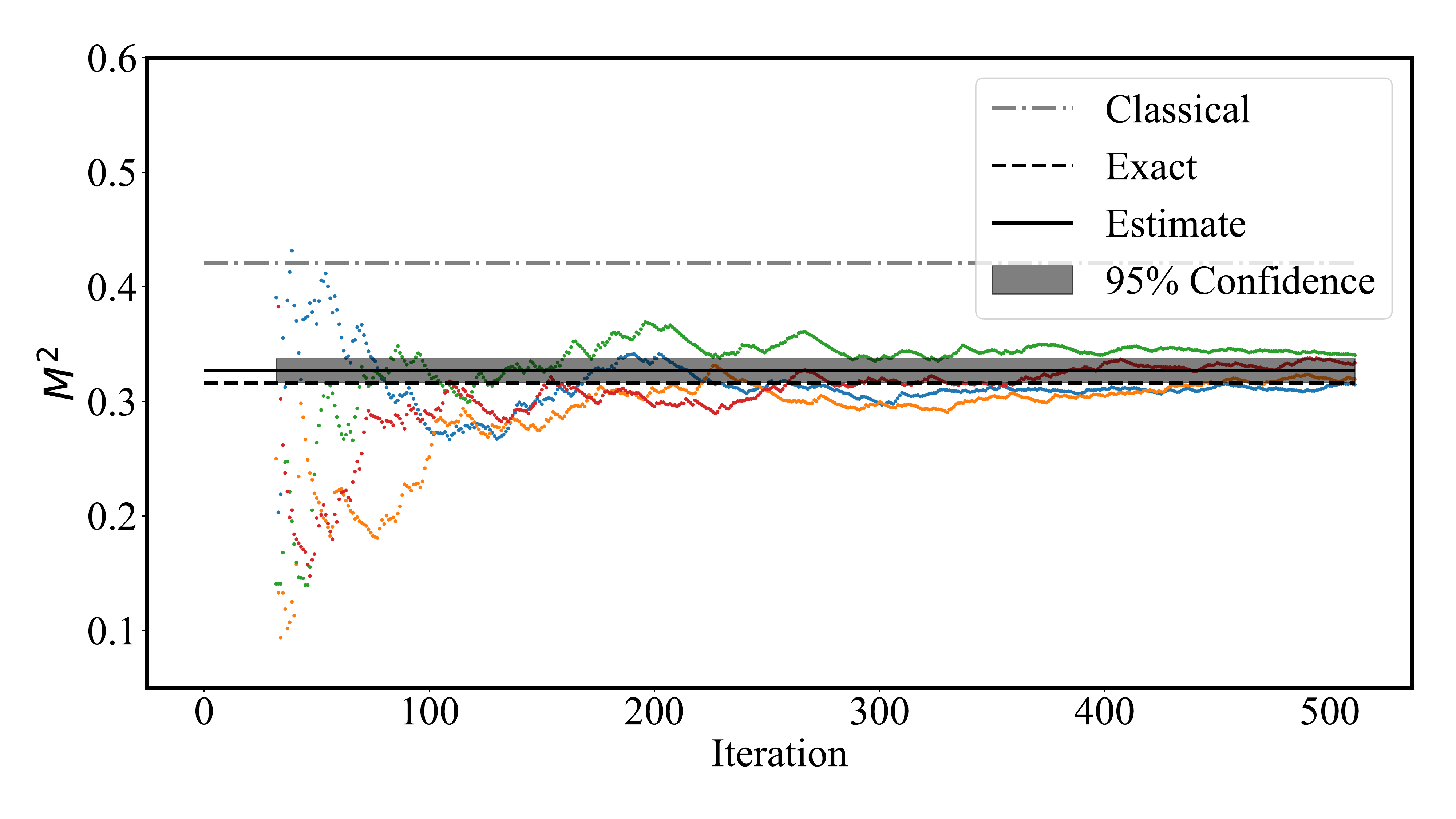}
	\caption{
		Evolution of average squared magnetization for four different Monte Carlo chains of the 16-site TFIM with $h_x=1$ and $\beta=\beta_c$ using noisy Loschmidt echos via AerSimulator.
  Each colored point represents the average of the chain up to the specified iteration.
	The first $32$ iterations (burn-in period) are excluded.
		The classical  refers to the value of the equivalent Ising model (i.e., with $h_x=0$).
	}
	\label{fig:chains}
\end{figure}

In Fig.~\ref{fig:chains}, we show how the average squared magnetization evolves with iterations of different Monte Carlo chains at $\beta_c$.
The exact result falls within two standard deviations from the final estimated average.
On the other hand, the classical result (i.e., for $h_x=0$) is well outside the error bars, so we could not have obtained these results by simply using the classical Boltzmann weights $e^{-\beta\braket{\psi |\hat{H}|\psi}}$.
This comparison is relevant because the quantile filter truncates the tails of the density of states, making it look closer to the density of states of the classical Ising model (which is a delta function at the mean energy). Therefore, in the presence of noise, the estimated quantum Boltzmann weights tend to get biased towards the classical Boltzmann weights, and the results here show that, at this temperature, the truncation does not significantly bias the estimates towards the classical model.

\begin{figure}[t]
	\centering
	\includegraphics[width=\columnwidth]{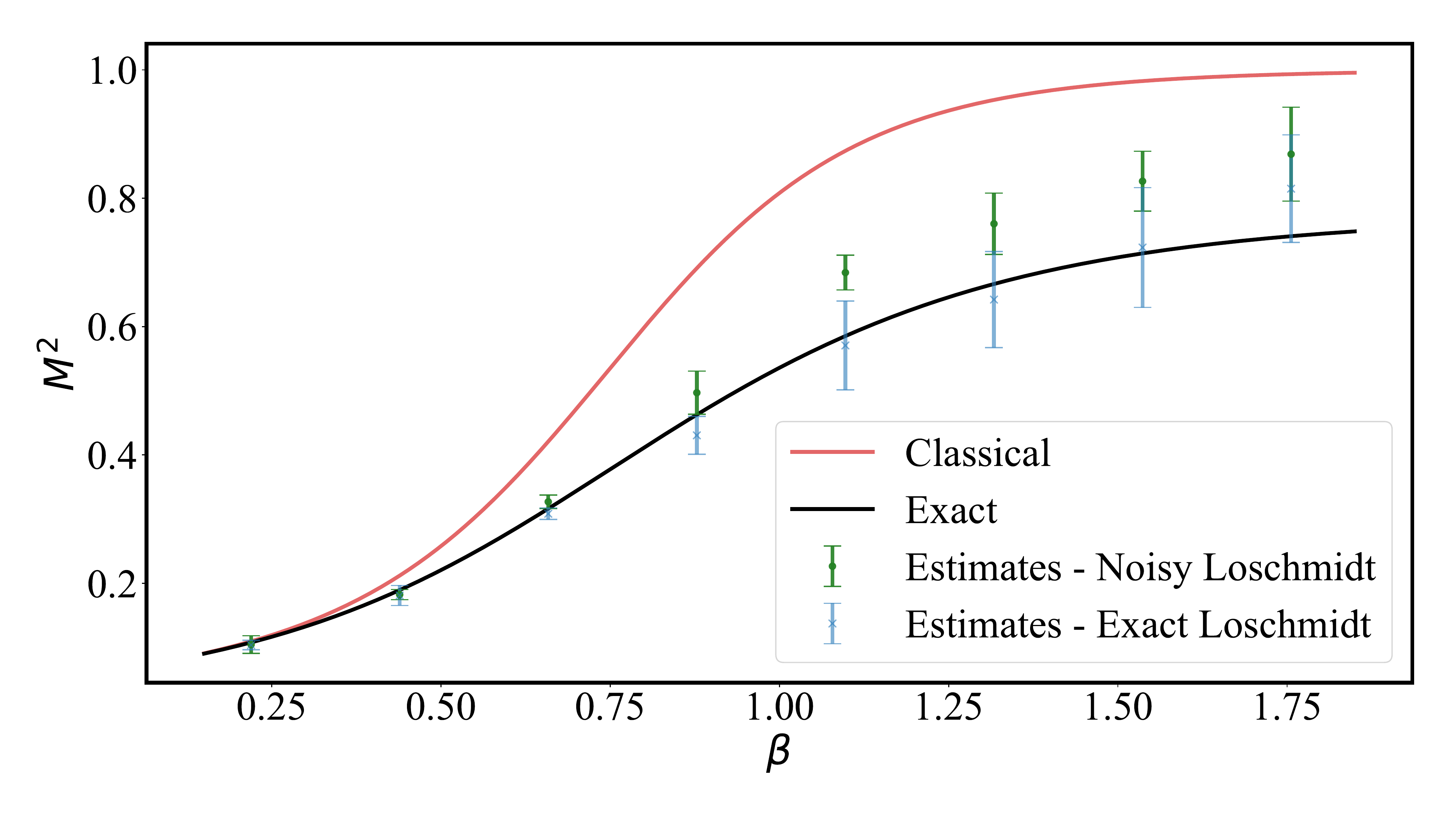}
	\caption{
		Average squared magnetization of the 16-site TFIM with $h_x=1$  at different inverse temperatures $\beta$. 
		Estimates refer to the simulation results of the time-series Monte Carlo algorithm using either noisy Loschmidt echos (via AerSimulator) or exact ones (via QuSpin).
		Error bars represent 95\% confidence (two standard deviations).	
		The classical curve represents the values for the equivalent Ising model (i.e., with $h_x=0$).
	}
	\label{fig:simulation_results}
\end{figure}

In Fig.~\ref{fig:simulation_results}, we compare the estimated values of squared magnetization using AerSimulator (plotted in green) with the exact ones (plotted in black) at inverse temperatures  $\beta_c/3, 2\beta_c/3, ..., 8\beta_c/3$.
We see that the bias is limited to 7.4\%  even below the critical point at inverse temperature $4\beta_c/3$.
As expected, the bias gets more prominent as the temperature decreases.
To disentangle the source of this bias, we also performed simulations using Loschmidt echos obtained numerically via QuSpin Python package.
These Loschmidt echos were calculated at the same time points as in the Aer simulations, but they are numerically exact, i.e., without Trotter error or shot noise.
Accordingly, the Boltzmann weights were estimated using NNLS without requiring the quantile filter.
The results using these exact Loschmidt echos (plotted in blue) show no observable bias beyond the statistical error bars. 
This indicates that the bias in AerSimulator results can be mainly attributed to Trotter error and the shot noise.
Reducing the bias involves taking more shots and smaller Trotter steps.
We also notice that statistical error bars get larger for lower temperatures. 
This is explained by the lower efficiency of the cluster update at lower temperatures, which has been discussed earlier at the end of Sec.~\ref{sec:monte_carlo}.
Reducing these error bars is a matter of taking more Monte Carlo samples.

\section{Summary and Outlook}
Time-series quantum Monte Carlo is a promising hybrid algorithm for calculating the thermal properties of quantum materials.
The algorithm relies on calculating quantum Boltzmann weights via Wick's rotation of real-time dynamics to imaginary time.
In this work, we studied in detail how to calculate Boltzmann weights from noisy and truncated time series.
We revisited the Gaussian filter method and discussed an automatic way of determining its cut parameter.
Under simplified assumptions, we derived an asymptotic of the error made on Boltzmann weights and found that lower temperatures require proportionally longer times to maintain the same level of accuracy.
We then proposed to use the non-negative least squares method, which enforces the non-negativity of the density of states. We found it to reduce the error in the Boltzmann weights by several orders of magnitudes compared to the Gaussian filter method using the same data.
Numerical results suggest that the duality between maximum time and temperature still holds using this method.

To regularize the effect of shot noise, we suggested using a quantile filter for truncating noisy tails of the density.
The amount of truncation is determined automatically via the discrepancy principle.
Benchmarking results show that the number of shots required to maintain a specific level of accuracy scales exponentially with inverse temperature.
It also suggests that entangled states with lower energy variances can be less susceptible to shot noise.
To test the robustness of the recipe in realistic settings, we fully implemented the time-series quantum Monte Carlo algorithm for the two-dimensional TFIM, including quantum circuits and an improved sampling strategy based on the cluster update algorithm.
The simulations indicate that, despite the aforementioned sensitivity to shot noise, the algorithm is able to obtain relatively accurate results up to the critical temperature using affordable resources.
Hardware errors aside, the parameters used in these simulations are within the capabilities of the current hardware.
This demonstrates that the method can be of practical use on current NISQ devices.

While the ferromagnetic TFIM model investigated here is relatively easy to solve on classical computers, the improvements made in Wick's rotation are general.
In particular, we expect the method to be useful for frustrated quantum spin models, e.g., the antiferromagnetic Heisenberg model on a kagome lattice, which is hard to simulate classically.
Also, the proposed reweighted cluster update can be similarly applied in that situation by proposing samples from an efficient classical sampling algorithm, e.g. the KBD algorithm~\cite{Kandel90, Zhang94}, and reweighting with the ratios of quantum to classical Boltzmann weights.
Whether the reweighted algorithm remains efficient at relevant low temperatures and large system sizes remains to be investigated.

\section{Acknowledgment}
The  authors  gratefully  acknowledge  funding  from  the German Federal Ministry of Education and Research (BMBF) under the EQUAHUMO project (grant number 13N16069). The authors thank Eleanor Crane, Kevin Hemery, Mohsin Iqbal and Ramil Nigmatullin for helpful discussions.
The authors also thank David Muñoz Ramo, Cono Di Paola and Luuk Coopmans for providing useful
feedback.

\appendix
\section{Error Scaling in Gaussian Filter Method}\label{app:error_scaling}
There are various sources of error in the estimation of Boltzmann weights in the Gaussian filter method.
First, there is an interference error resulting from the finite sampling frequency $\pi/\Delta t$ because the broadened density has infinite support (see Fig.~\ref{fig:interference}).
There is also a quadrature error associated with performing the integral numerically with a finite number of $\omega$ points.
These two sources of error are controlled by using finer time and frequency grids, and we assume that  
$\Delta t$ and $\Delta \omega$ are chosen small enough to make the resulting errors negligible.
More importantly, there is an error introduced by cutting values below $D_\psi^\text{cut}$ and errors on the other values remaining from having a finite maximum time $T_\text{max}$.
These errors are dominated by the error resulting from setting the values at the lowest frequencies to zero.
Let $\omega_\text{cutoff}$ be the smallest frequency below which all values of the density are truncated.
That error can then be calculated as
\begin{equation}
	E_{\psi, \delta}^\text{cut} = \int_{-\infty}^{\omega_\text{cut}} d\omega \ e^{-\beta \omega} D_{\psi, \delta}(\omega).
\end{equation}

To simplify the analysis, we focus on the case where the exact density is a single delta function at frequency $\omega_0$ and discuss implications for the general case later.
In this simple case, the broadened density is a Gaussian and the exact value of its Boltzmann weight is  $W_{\psi, \delta} = \exp(\beta^2 \delta^2/2 - \beta \omega_0)$. The relative error can  then expressed in terms of the error function as following:
\begin{equation}\label{eq:rel_err}
	E^\text{cut}_{\psi, \delta}/W_{\psi, \delta} = \frac{1}{2} \left[1 + \operatorname{erf} \left(\frac{\beta \delta^2 + \omega_\text{cut}-\omega_0}{\sqrt{2} \delta}\right)\right]
\end{equation}
The distance between the peak and cutoff frequency $\omega_\text{cut}-\omega_0$ will depend on the maximum time $T_\text{max}$ and broadening parameter $\delta$ and can be estimated as follows.
The density of the truncated time series reads
\begin{align}
		D_{\psi, \delta}^{T_\text{max}}(\omega) &= \frac{1}{2\pi}\int_{-T_\text{max}}^{+T_\text{max}} dt \ e^{- t^2 \delta^2/2}\  e^{it (\omega-\omega_0)} \nonumber \\
		 &= \frac{i e^{-a^2}}{2 \sqrt{2\pi} \delta} \left[\operatorname{erfi}(a-ib) - \operatorname{erfi}(a+ib)\right]
\end{align}
where $\operatorname{erfi}$ is the imaginary error function and we defined $a \coloneqq (\omega-\omega_0)/(\sqrt{2} \delta)$ and $b \coloneqq \delta\ T_\text{max}/\sqrt{2}$ for compactness.
In the limit $b \propto \delta\ T_\text{max} \to \infty$, we can use the expansion of $\operatorname{erfi}(z)$ at positive and negative infinities to order $\mathcal{O}(z^{-3} e^{z^2} )$ and write
\begin{equation}
	D_{\psi, \delta}^{T_\text{max}}(\omega) \approx \frac{e^{-a^2}}{\sqrt{2\pi} \delta} \left[ 1 + e^{a^2-b^2} \frac{ a \sin(2 a b) - b \cos(2 a b)}{\sqrt{\pi}  \ (a^2+b^2)} \right]
\end{equation}
The first term  gives  the exact density of the full time series, while the second one approximates the error made due to the time-series truncation.  
The oscillations start to dominate when the  exponential factors $e^{a^2}$ and $e^{b^2}$  have similar magnitudes. 
This allows us to infer how the frequency cut $\omega_\text{cut} $ scales
\begin{equation}
	a^2 = C b^2 \Rightarrow \omega_\text{cut} - \omega_0 =   - C \delta^2 T_\text{max}
\end{equation}
where $C$ is some positive constant of order 1.
Substituting back in Eq.~\eqref{eq:rel_err}, the relative error reads
\begin{equation}
	E^\text{cut}_{\delta}/W_{\psi, \delta} = \frac{1}{2} \left[1 + \operatorname{erf} \left(\frac{\delta\ T_\text{max}}{\sqrt{2}} \left[ \frac{\beta}{T_\text{max}}   - C\right] \right)\right]
\end{equation}
In the same limit as above, $\delta\ T_\text{max} \to \infty$, and assuming $\beta < C T_\text{max}$, we can expand the error function $\operatorname{erf}(z)$ at negative infinity to order $\mathcal{O} (z^{-3} e^{-z^2})$ and approximate the relative error as
\begin{equation}
	E^\text{cut}_{\delta}/W_{\psi, \delta} \approx \frac{\exp \left(-\delta^2\ T_\text{max}^2\left[ \beta/T_\text{max}  - C\right]^2/2 \right)}{\sqrt{2 \pi}\ \delta\ T_\text{max}  \left[C - \beta/T_\text{max}  \right]}.
\end{equation}
Setting $\alpha \coloneqq \delta\ T_\text{max}$, Eq.~\eqref{eq:error_scaling} is obtained.

Now we discuss the general case of a linear combination of delta functions.
Let $\mu$ be the mean of the density and $\sigma$ is its standard deviation.
Assuming $\delta$ is much larger than the spacing between the lowest and highest frequencies, for the sake of error scaling, the broadened density could be approximated by a Gaussian of width $\sigma+\delta$ and mean $\mu$.
With this assumption, the previous error analysis holds by setting $\omega_0$ to the mean $\mu$ and replacing $\delta$ by the extended width $\sigma+\delta$.

\bibliography{manuscript}

\end{document}